\documentclass[]{pasj01}
\draft

\Received{11 February 2019}
\Accepted{12 May 2019}
 
\usepackage{multirow}
\usepackage{flexisym}
\usepackage{lscape}
\usepackage{graphicx}
\usepackage{epstopdf}
\usepackage{color}
\usepackage[font=normalsize]{caption}
  
\begin{document} 

\title{ 
	{}Large angular scale fluctuations of near infrared extragalactic background light based on the IRTS observations }

\author{Min Gyu \textsc{Kim}\altaffilmark{1}}%
\email{mgkim@astro.snu.ac.kr}

\author{Toshio \textsc{Matsumoto},\altaffilmark{2}}

\author{Hyung Mok \textsc{Lee}\altaffilmark{1,3}}

\author{Woong-Seob \textsc{Jeong}\altaffilmark{3,4}}

\author{Kohji \textsc{Tsumura}\altaffilmark{5}}

\author{Hyunjong \textsc{Seo}\altaffilmark{3}}

\author{Masahiro \textsc{Tanaka}\altaffilmark{6}}

\altaffiltext{1}{Department of Physics and Astronomy, Seoul National University, Seoul 08826, Korea}
\altaffiltext{2}{Department of Space Astronomy and Astrophysics, Institute of Space and Astronautical Science (ISAS), Japan Aerospace Exploration Agency (JAXA), 3-1-1 Yoshinodai, Chuo-ku, Sagamihara, Kanagawa 252-5210, Japan}
\altaffiltext{3}{Korea Astronomy and Space Science Institute (KASI), 776 Daedeok-daero, Yuseong-gu, Daejeon 34055, Korea}
\altaffiltext{4}{University of Science and Technology, Korea (UST), 217 Gajeong-ro, Yuseong-gu, Daejeon 34113, Korea}
\altaffiltext{5}{Department of Natural Science, Faculty of Knowledge Engineering, Tokyo City University, 1-28-1, Tamazutsumi, Setagaya, Tokyo, 158-8557, Japan}
\altaffiltext{6}{Center for Computational Sciences, University of Tsukuba, 1-1-1 Tennodai, Tsukuba, Ibaraki 305-8577, Japan}


\maketitle

\clearpage

\begin{abstract}
	We measure the spatial fluctuations of the Near-Infrared Extragalactic Background Light (NIREBL) from 2$^{\circ}$ to 20$^{\circ}$ in angular scale at the 1.6 and 2.2 $\micron$ using data obtained with Near-Infrared Spectrometer (NIRS) on board the Infrared Telescope in Space (IRTS). The brightness of the NIREBL is estimated by subtracting foreground components such as zodiacal light, diffuse Galactic light, and integrated star light from the observed sky. The foreground components  are estimated using well-established models and archive data. The NIREBL fluctuations for the 1.6 and 2.2 $\micron$ connect well toward the sub-degree scale measurements from previous studies. Overall, the fluctuations show a wide bump with a center at around 1$^{\circ}$ and the power decreases toward larger angular scales with nearly a single power-law spectrum (i.e. \textit{F($\sqrt{l(l+1)C_l/2\pi}$)} $\sim$ $\theta^{-1}$) indicating that the large scale power is dominated by the random spatial distribution of the sources. After examining several known sources, contributors such as normal galaxies, high redshift objects, intra-halo light, and far-IR cosmic background, we conclude that the excess fluctuation at around the 1$^{\circ}$ scale cannot be explained by any of them.
\end{abstract}

\newpage

\section{INTRODUCTION}

	The Near-Infrared Extragalactic Background Light (NIREBL) is the integrated light of the entire cosmic history in the near infrared. Thus, the origin of the NIREBL is essential to probe the formation and evolution of galaxies from birth to the present Universe. Since current technology limits us from resolving diffuse, faint, and distant objects that contribute to the NIREBL brightness, we should rely on measurements of spatial fluctuations and absolute brightness to understand the nature of the NIREBL. The absolute brightness measures the background intensity, and the spatial fluctuation measures the clustering properties of the emitting sources.
	
	The first reliable measurement of the absolute brightness conducted at 1.25, 2.2, 3.5, and 4.9 $\micron$ with the Diffuse InfraRed Background Experiment (DIRBE) on the Cosmic Background Explorer (COBE) although it experienced difficulties in subtracting the contribution from Galactic stars due to large confusion limit \citep{gorjian00,wright00,cambresy01,levenson07,sano15,sano16}. They found 2 to 8 times larger brightness than the Integrated Light of Galaxies (ILG). Thanks to the smaller beam size and low resolution spectrograph, the Infrared Telescope in Space (IRTS) confirmed the NIREBL excess by observing the isotropic background spectrum at short wavelengths (1.4 - 4 $\micron$) with better precision \citep{matsumoto05,matsumoto15}. With better point source subtractions, AKARI also succeeded to confirm the excess NIREBL spectrum at 2 - 5 $\micron$  \citep{tsumura13b}. The spectra obtained by COBE, IRTS, and AKARI are consistent within the common wavelength region ($\lambda$ $>$ 2 $\micron$). Several Extragalactic Background Light (EBL) observations were also carried out at around the optical wavelength range \citep{bernstein07,matsuoka11,mattila17a,mattila17b,kawara17,matsuura17,zemcov17}. However, their brightness levels are not in good agreement at $<$ 0.7 $\micron$, and this discrepancy may have been caused by uncertainties in the foregrounds subtraction.
	
	The excess brightness was initially explained by the first generation of stars that formed at the reionization era \citep{santos02,salvaterra03}. However, theoretical models based on the recent observations of high redshift galaxies indicate that the first stars contribute less than 1\% of the total absolute flux of the observed EBL \citep{cooray12b,yue13}.
	On the other hand, several studies argue that the excess brightness is not a real background but a measurement error. For example, \citet{dwek05} and \citet{kawara17} tried to explain the excess brightness with a subtraction error of the Zodiacal Light (ZL) which is the brightest diffuse foreground component.
	
	Unlike the absolute brightness measurement, the spatial fluctuation can be measured to mitigate the problem of foreground subtraction since the fluctuation is less sensitive to a foreground component. For example, although the ZL is the brightest foreground component, it is expected that ZL is very smooth over the large angular scales \citep{abraham97,pyo12}. Therefore, the EBL fluctuation can be more clearly distinguished from ZL. The detection of an excess EBL fluctuation was measured by \citet{kashlinsky05} with Spitzer at angular scales up to 5$\arcmin$ in wavelengths between 3.6 to 8 $\micron$, after subtracting the contribution from galaxies brighter than mag$_{AB}$ = 25. Subsequently, an excess fluctuation was detected over the ILG at angular scales up to 1$^{\circ}$, confirming the previous measurements \citep{kashlinsky12,cooray12a,mitchell15} using deeper and wider data from Spitzer. Using the HST data, \citet{thompson07} and \citet{donnerstein15} measured fluctuations at 1.1 and 1.6 $\micron$ at the sub-arcminute scale. \citet{matsumoto11} detected excess fluctuation above 100$\arcsec$ from AKARI (2.4, 3.2, and 4.1 $\micron$) data and found that the fluctuation follows Rayleigh-Jeans like spectrum (i.e. $\lambda$I$_\lambda$ $\sim$ $\lambda^{-3}$). Using a wider field AKARI image, \citet{seo15} found the existence of excess power up to $\sim$ 0.3$^{\circ}$. At 1.1 and 1.6 $\micron$, the Cosmic Infrared Background ExpeRiment (CIBER) measured large angular scale ($<$ 1$^{\circ}$) fluctuations \citep{zemcov14}. They reported a clear excess fluctuation at angular scales between 0.1$^{\circ}$ and 0.36$^{\circ}$. They used the Intra Halo Light (IHL) at z $<$ 3 to explain the excess fluctuation \citep{cooray12a,zemcov14}. The IHL source consists of tidally stripped stars during galaxy mergers and interactions. However, the IHL is not observationally confirmed and cannot explain all of the observed excess fluctuation. Consequently, there is no clear consensus regarding the origin of the NIREBL.
	
	To understand the origin of the excess fluctuations, we examine the fluctuation spectrum using IRTS data with a scale up to several-degrees. Such large scale fluctuations has never been explored before. Our approach can also constrain the physical properties of the excess origins. The outline of this paper is as follows. In section \ref{S:instrument}, we introduce the instrument. We describe the observation and data reduction in section \ref{S:irtsobs}. The data analysis is described in section \ref{S:datareduction}. The power spectrum estimation and the result are shown in section \ref{S:irtsps} and \ref{S:result}, respectively. In section \ref{S:discussion}, the discussions are given. Finally, we summarize our result in section \ref{S:summary}.
	
	\section{INSTRUMENT}
	\label{S:instrument}

	IRTS, the first Japanese orbiting IR telescope onboard the Space Flyer Unit (SFU), was launched on March 18 UT in 1995. It surveyed 7\% of the sky until its liquid Helium was exhausted on April 25. The IRTS is a 15 cm Ritch-Chretien type telescope with a focal length of 60 cm. The whole system, together with four focal plane instruments, was cooled down to 2 K using liquid Helium \citep{murakami96}. 
	
	Among those instruments, the Near-Infrared Spectrometer (NIRS) is optimized to study the diffuse background with deep and wide sky coverage. The NIRS covers the wavelength range between 1.4 and 4.0 $\micron$ with a 0.13 $\micron$ spectral resolution. The incident beam goes through a 1.4 mm $\times$ 1.4 mm slit which corresponds to an 8$\arcmin$ $\times$ 8$\arcmin$ area in the sky, and it is diffracted by the grating. The dispersed beam is then focused on the linear array consisting of 24 InSb detector elements \citep{noda94}. To reduce the background errors arising from Galactic stars, it has a higher spatial resolution than DIRBE, and the cold shutter is installed to obtain a dark current. The stability of the detector is monitored using a calibration lamp during the observation. It uses J-FET charge integrating amplifiers operating at 60 K to detect a low background brightness by reducing the noise and achieving a high sensitivity. The InSb detector reads outs data with a 4 Hz sampling rate. The charges were integrated for 65.54 second before a reset and 8.192 seconds of it was used for dark current observation with a shutter close configuration. Details of the NIRS performance is found in \citet{noda96}.
	
	\section{OBSERVATION AND DATA REDUCTION}\label{S:irtsobs}

	Of the entire IRTS coverage, we used data initially reduced by \citet{matsumoto05}. They used data obtained at Galactic latitudes above 40$^{\circ}$ to avoid the strong foreground emissions due to the stars and dust in the Galaxy. The data obtained, while passing through the South Atlantic anomaly region where the noise level increases with high energy charged particles, was rejected. Of the 65.54 second charge integration between resets, first 4 second data due to anomalous residual charges after the reset was not used. The flux(e$^{-}$ s$^{-1}$) of each IRTS data was then obtained from linear fit of charges for 5 seconds along the scan direction. In linear fit process, contaminated data by cosmic-rays, instrumental noise, and stars were excluded. Dark current was subtracted after the linear fit process. Details of this process is described in \citet{matsumoto05}.

	Astrometry was achieved within 2.2$\arcmin$ using an attitude control sensor that was accurate enough to identify the bright Galactic stars \citep{murakami96}. The absolute calibration was achieved with a few percent errors using the standard stars observed by the IRTS \citep{noda96}. The calibration factor measured from the laboratory and that derived from the observed stellar fluxes were in good agreement. The final data at 24 discrete bands covers 1\% of whole sky. From these IRTS spectra, we made the synthesized 1.6 and 2.2 $\micron$ band fluxes. Specifically, fluxes from 1.53, 1.63, and 1.73 $\micron$ were averaged to obtain the 1.6 $\micron$ flux and those from 2.03, 2.14, 2.24, and 2.34 $\micron$ were averaged to obtain the 2.2 $\micron$ flux (hereafter, IRTS SKY).
	
	\section{DATA ANALYSIS}
	\label{S:datareduction}
	
	To measure the spatial fluctuation of the NIREBL, we need the brightness map of the background. The background brightness can be derived by subtracting brightness of all astrophysical foreground components from the observed sky brightness. In this section, we describe how we estimate brightness of the observed sky and foregrounds such as Diffuse Galactic Light (DGL), Integrated Star Light (ISL), and ZL.
	
	We also performed the pixelization for the estimated brightness of each component. They were pixelized into pixels covering a nearly equal area for the power spectrum analysis since the IRTS unevenly scanned the sky. To do this, we used well-developed tool HEALPix, which stands for the Hierarchical Equal Area isoLatitude Pixelization \citep{gorski05}. HEALPix divides the surface of the sphere into pixels of roughly equal shape and identical size. The resolution of the pixelization is defined by N$_{side}$ = 2$^{k}$, where $k$ can be any positive integer. The number of pixels in the whole sky is 12 N$_{side}^2$. By considering the IRTS FoV (i.e. 20$\arcmin$ $\times$ 8$\arcmin$), we used N$_{side}$ $=$ 64, which corresponds to a 55$\arcmin$ $\times$ 55$\arcmin$ pixel size. This divides the whole sky into 49152 pixels.

	\subsection{IRTS data analysis}

\begin{figure}
		\includegraphics[width=160mm]{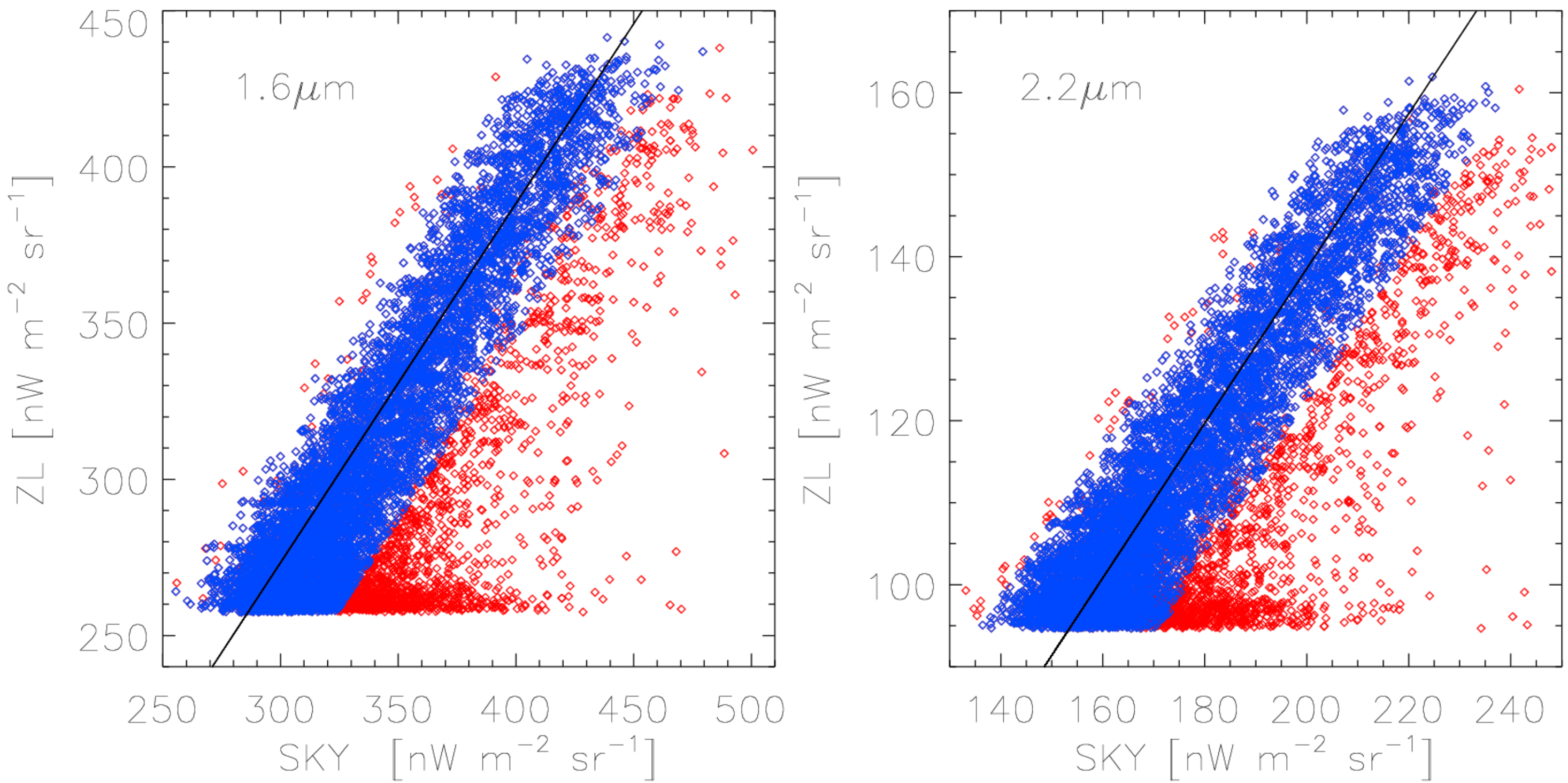}
		\caption{%
		The brightness correlation between the IRTS SKY and IRTS ZL. Left and right panels are 1.6 and 2.2 $\micron$, respectively. Red symbol indicates raw data before clipping process. Blue symbol indicates remained data after the clipping process. Black solid line is linear fit along the most dense data regions in the raw data.}%
		\label{F0}
	\end{figure}

	In this section, we describe additional clipping process which was performed before the pixelization of the IRTS SKY. The clipping process is as follows. Using the IRTS SKY described in section \ref{S:irtsobs}, we did the correlation analysis with the ZL (see figure \ref{F0}). Here, the estimation of the ZL is described in section \ref{SSS:irtszl}.
	
\begin{figure}
		\includegraphics[width=160mm]{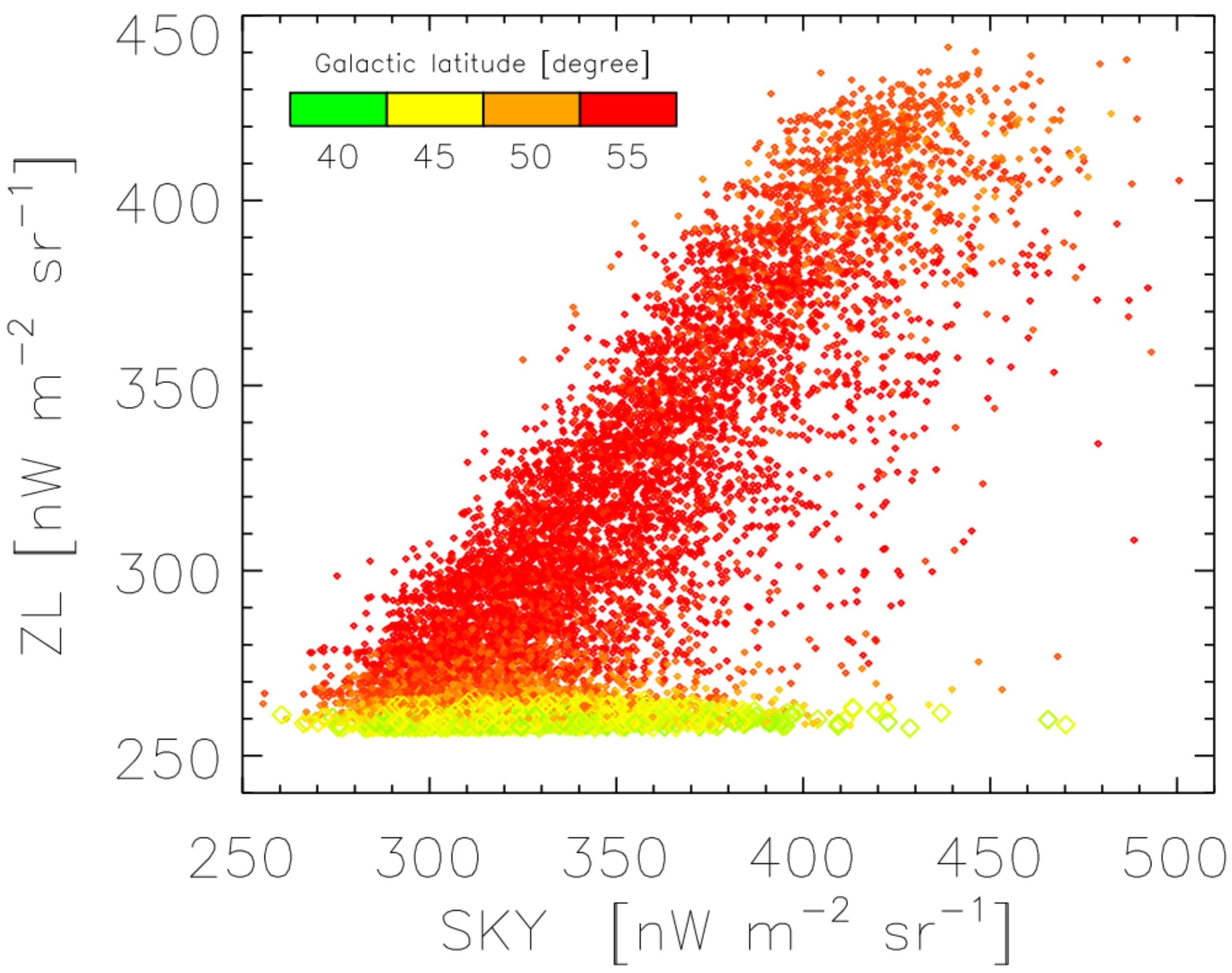}
		\caption{%
		Same correlation diagram as shown in figure \ref{F0} at 1.6 $\micron$. Each data is colored by its Galactic latitude as indicated by color bar. To highlight the data at low Galactic latitude region (\textit{b} $<$ 45$^{\circ}$), we draw them with larger symbol than others.}%
		\label{F15}
	\end{figure}

	Since the ZL brightness accounts for more than 90\% of the IRTS SKY, figure \ref{F0} should show strong correlation. However, some of the IRTS data located at low ZL brightness region does not follow the correlation. These outliers are mostly located at lower Galactic latitude region as shown in figure \ref{F15}. This indicates that they are due to residual bright stars which was not rejected from the data reduction process. To reject these data, we linearly fitted the correlation diagram in figure \ref{F0} and excluded the IRTS data out of 2$\sigma$ which is around two times larger than the brightness range of the ISL. The fitting process was done as follows. On the correlation diagram in figure \ref{F0}, we divided the y-axis (i.e. ZL) into constant brightness interval. At each interval, we determined the most dense data region along the x-axis (i.e. IRTS SKY). Finally, the linear fit was done along those most dense data regions. Around 18 \% of the IRTS data was rejected through this process.
	
\begin{figure}
		\begin{center}
		\includegraphics[width=160mm]{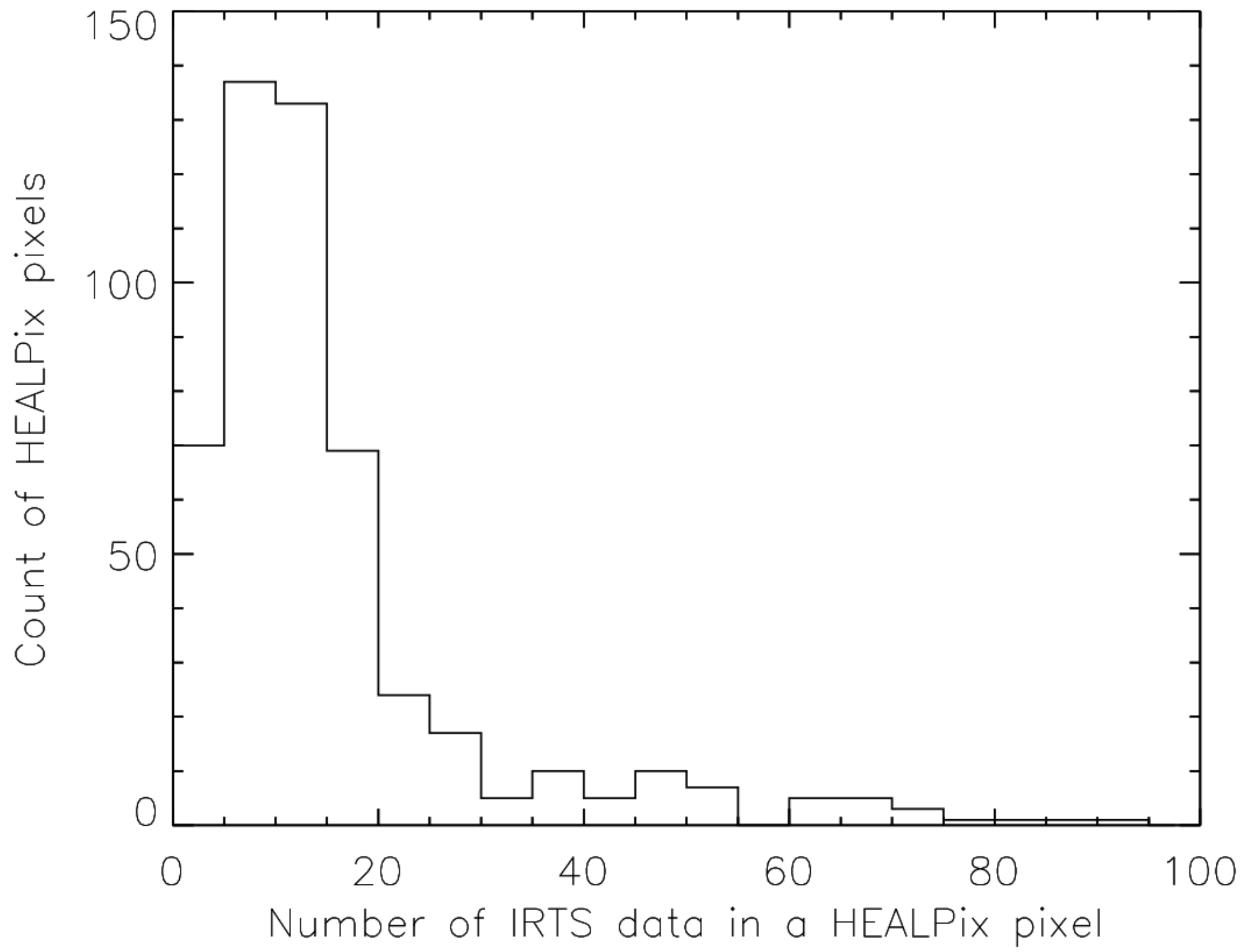}
		\end{center}
		\caption{%
		Number of the IRTS data belong to each HEALPix pixel. The brightness of a HEALPix pixel is mean from the belonging IRTS data. Here, the bin size of the histogram is 5.}%
		\label{F1}
	\end{figure}

	The residual IRTS data for each 1.6 and 2.2 $\micron$ was then assigned to the HEALPix pixels according to their positions. Due to the large HEALPix pixel size, a few to tens of the IRTS data belong to each HEALPix pixel. The number of IRTS data in a HEALPix pixel is shown in figure \ref{F1}. The intensity of each HEALPix pixel was assigned from a mean brightness of the IRTS SKY in the HEALPix pixel. If the number of IRTS fields in a HEALPix pixel is less than 5, we masked the HEALPix pixel. They are 14\% of the HEALPix pixels covered by the IRTS fields. The pixelized 1.6 and 2.2 $\micron$ maps were stored in the FITS format \citep{calabretta02} for the next step to evaluate the power spectrum of the IRTS SKY.
	
	\subsection{Foregrounds data analysis}
	
	In this section, we describe how we estimated the foreground brightness such as DGL, ISL, and ZL. Each foreground brightness was then pixelized into HEALPix scheme.
	
	\subsubsection{Diffuse Galactic Light}\label{SSS:irtsdgl}

\begin{figure}
		\includegraphics[width=160mm]{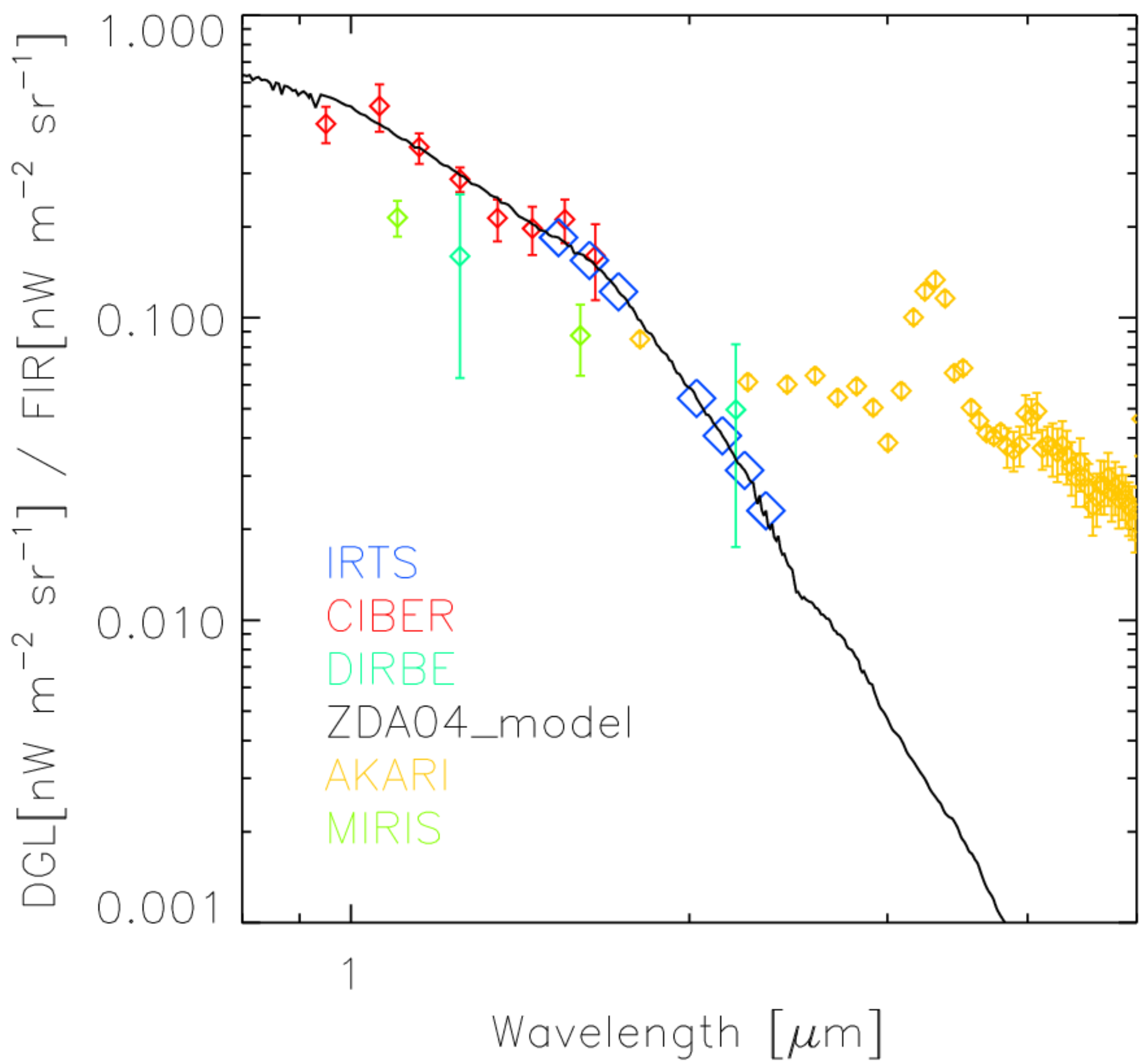}
		\caption{%
		DGL spectrum normalized by far-IR emission at 100 $\micron$. The CIBER/LRS is from \citet{arai15}, and DGL model to fit the CIBER/LRS data is drawn with solid line (ZDA04; \cite{brandt12}). Blue diamond symbols are points for the IRTS bands. The DIRBE is from \citet{sano15}, AKARI is from \citet{tsumura13a}, and MIRIS is from \citet{onishi18}. Here, the AKARI data is mainly contributed by the PAH emission. Since the IRTS sky coverage is far from the Galactic plane (b $>$ 40$^{\circ}$) where contributions of PAH emission is negligible, we did not consider the AKARI data.}%
		\label{F2}
	\end{figure}

	The DGL consists of star light scattered from dust grains distributed in interstellar space. Since the DGL is diffuse and faint, it is difficult to observe directly. Nevertheless, far-IR thermal emission (e.g. 100 $\micron$), HI or CO column density \citep{brandt12} has close relation with the DGL brightness. At near-IR wavelength region, the relation between the 100 $\micron$ thermal emission and the DGL brightness has been studied based on the near-IR observed data \citep{arai15, sano15}. They derived scale factor between the 100 $\micron$ thermal emission and the near-IR DGL brightness. Then, they fitted the various model spectra to the scale factor. The best fitted model was \citet{brandt12} model as shown in figure \ref{F2}. The fitted scale factor enables us to derive the near-IR DGL brightness from the 100 $\micron$ intensity.

	To estimate the DGL at IRTS fields using the scale factor as shown in figure \ref{F2}, we used the 100 $\micron$ thermal emission from the SFD dust map \citep{schlegel98}. The SFD dust map was also used for the scale factor derivation \citep{arai15, sano15}. Nevertheless, since the SFD map was not corrected for the cosmic infrared background brightness, we subtracted 0.8 MJy sr$^{-1}$ \citep{puget96,fix98,lagache00,matsuoka11} so the map contains only a dust emission component. From the brightness corrected SFD map, we obtained the 100 $\micron$ intensities belong to each IRTS FoV. 100 $\micron$ intensities at each IRTS FoV were then averaged. The averaged intensity was then multiplied by the scale factor at IRTS bands to derive the DGL brightness. The DGL brightness for the IRTS bands (i.e. 1.53, 1.63, 1.73, 2.03, 2.14, 2.24, and 2.34 $\micron$) was then averaged to make synthesized DGL brightness at 1.6 and 2.2 $\micron$ bands (hereafter, IRTS DGL). This process has been done for all IRTS fields and pixelized into the HEALPix scheme.

	\subsubsection{Integrated Star Light}\label{SSS:irtsisl}
\begin{figure}
		\includegraphics[width=160mm]{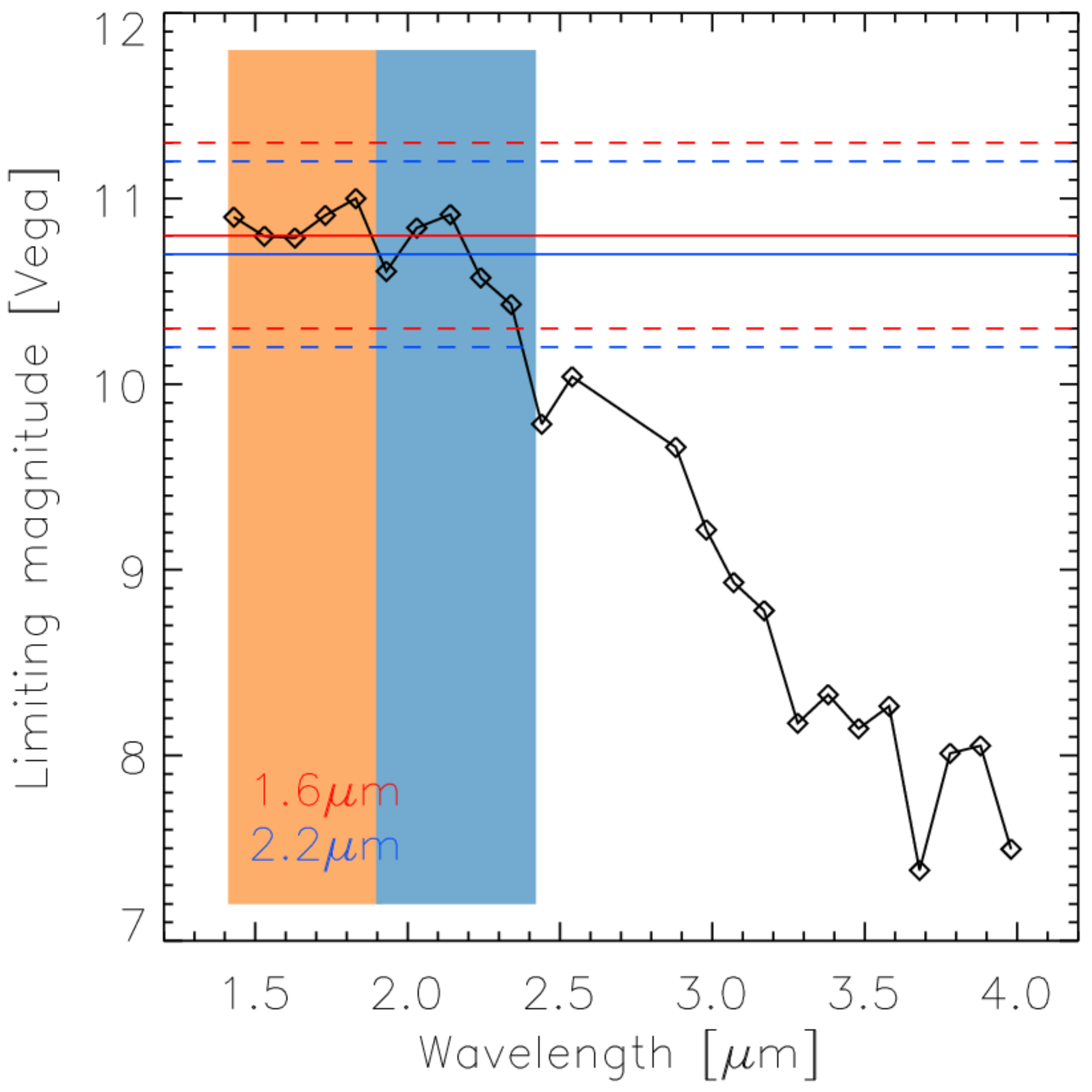}
		\caption{%
		Limiting magnitudes for the 24 IRTS bands (diamond symbol). Red and blue solid lines are the 1.6 and 2.2 $\micron$ limiting magnitudes used to estimate the brightness due to unresolved Galactic stars. Also drawn in dashed lines are $\pm$ 0.5 of limiting magnitude errors. Shaded colors represent bandwidths of 1.6 and 2.2 $\micron$.}%
		\label{F3}
	\end{figure}

	The ISL indicates the Galactic star light which contributes to the brightness of the observed IRTS SKY. We could remove the contributions of the bright stars from the observed sky brightness. However, the contributions of faint stars could not been removed. These faint stars are defined by limiting magnitude of the IRTS as shown in figure \ref{F3}. In this section, we describe how we estimated the contributions of faint stars.
	
	To estimate the ISL by faint stars above the IRTS limiting magnitude, we used a 2MASS point/extended source \citep{cohen03} which is well known Galactic star catalog at near-IR region. Since the limiting magnitude of the 2MASS star catalog is much fainter than that of the IRTS, we could estimate the ISL caused by stars lying between the IRTS and the 2MASS limiting magnitudes. To account the faint stars even above the 2MASS limiting magnitudes, we used Galactic model stars. The 2MASS limiting magnitudes for H- and K-bands are 15.1 and 14.3, respectively.
	
\begin{figure}
		\includegraphics[width=160mm]{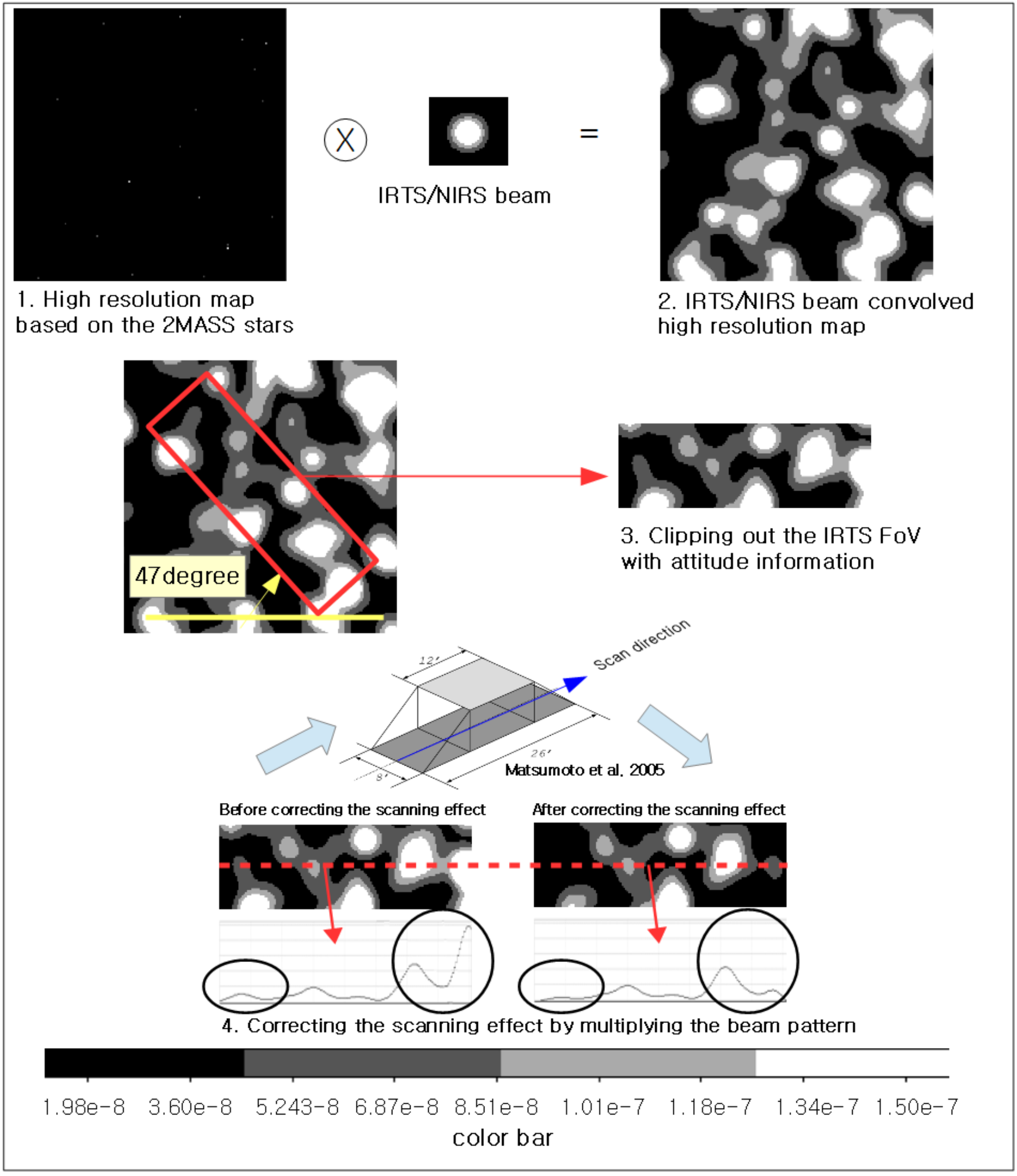}
		\caption{%
		Flow chart of the process to estimate the ISL brightness of an IRTS field based on the 2MASS stars.}%
		\label{Fislmap}
	\end{figure}

	First, we describe how we estimated the ISL for IRTS fields based on the 2MASS stars. The diagram of process flow to estimate the ISL for each IRTS field is shown in figure \ref{Fislmap}. That is, we reconstructed a high resolution map (pixel size is 10$\arcsec$) where the map size is more than twice larger than the IRTS FoV. Then, we distributed the 2MASS stars on the high resolution map where the center of the map is set by an nominal IRTS field position from the IRTS attitude information. Here, we used 2MASS stars lying between the IRTS and the 2MASS limiting magnitudes. Then, we convolved the high resolution map with the IRTS/NIRS beam pattern. The Full Width Half Maximum (FWHM) of the IRTS/NIRS beam is 2.4$\arcmin$. On the IRTS/NIRS beam convolved map, we clipped out regions covering the IRTS FoV with a center as same as IRTS nominal position. However, the scanning effect during the IRTS observation causing the beam pattern is not constant along the scan direction. The effective beam pattern is trapezoidal as shown in figure 2 of \citet{matsumoto05}. To apply this on the clipped map, we multiplied a factor showing trapezoidal pattern having unity for 12$\arcmin$ central region along the scan direction and linearly decreases to zero at both ends. This process has been done for all IRTS fields so the ISL based on the 2MASS stars was made.
	
	Next, to estimate the ISL by stars above the 2MASS limiting magnitude, we used the Galactic model stars named TRILEGAL \citep{girardi05}. The model provides bright to faint-end stars up to limiting magnitudes of 30. It models the number of stars and their brightness in various wavelengths toward any line of sight at a specified FoV. However, since the model star does not have astrometry information, we randomly assigned the astrometry information to model stars and distributed them on the high resolution map. Here, we used model stars fainter than the 2MASS limiting magnitude. Then, the ISL based on the model was estimated in the same manner as ISL derivation based on the 2MASS stars. However, the brightness of the model stars is known to have systematic offset comparing with that of the 2MASS stars (see figure 1 in \cite{matsumoto15}). To correct this, we multiplied factor of 1.23 to the brightness of the ISL based on the model stars. The ISL brightness for the 1.6 and 2.2 $\micron$ based on the model stars are 15\% and 24\% of the ISL brightness based on the 2MASS stars, respectively. Finally, total ISL (hereafter, IRTS ISL) for the IRTS fields was made by adding the ISL based on the 2MASS and the model. Then, the IRTS ISL was pixelized into the HEALPix scheme.
	
\begin{figure}
		\includegraphics[width=160mm]{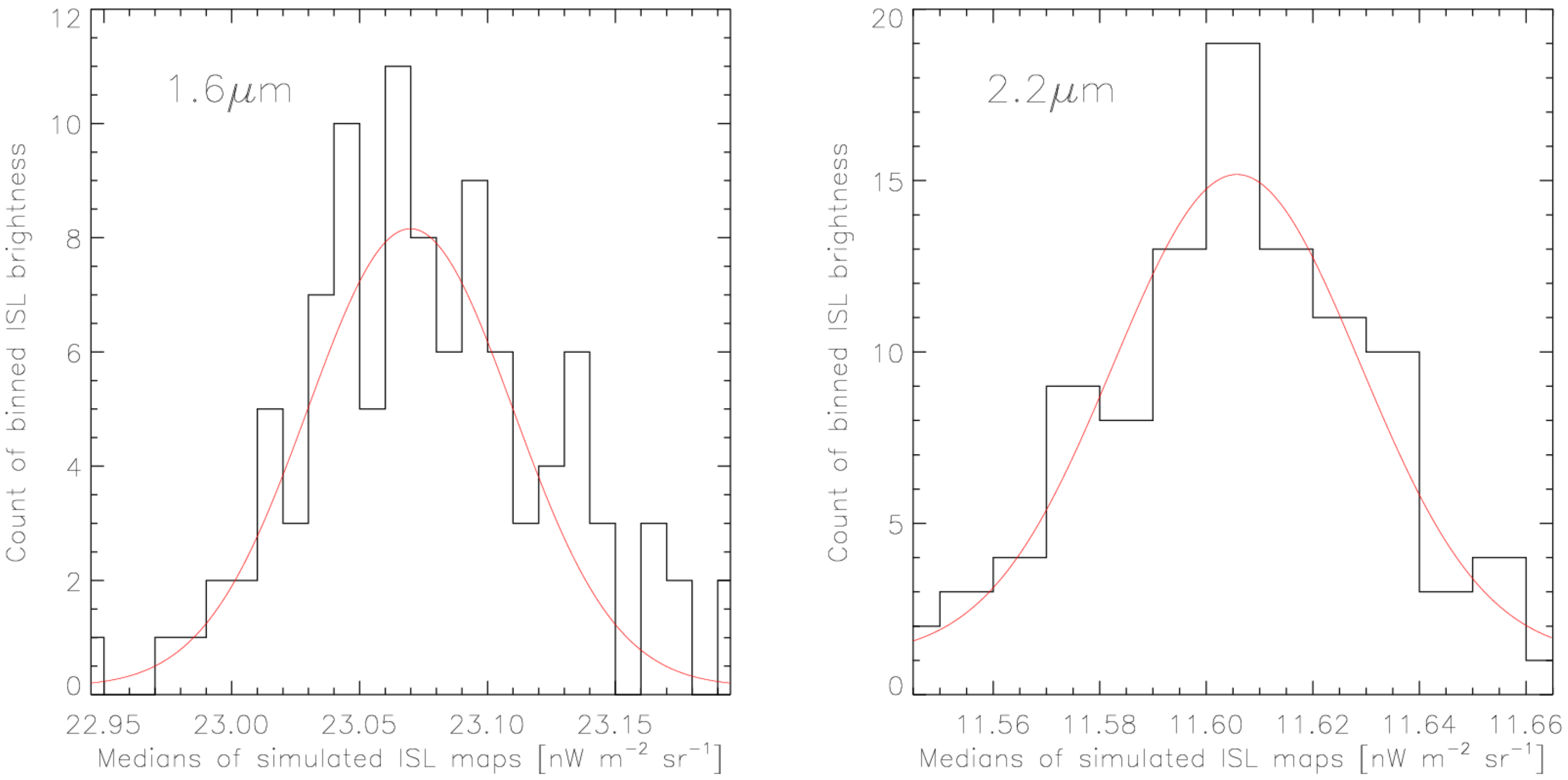}
		\caption{%
		Histogram of ISL medians for 100 simulated maps. Left and right panels are 1.6 and 2.2 $\micron$, respectively. Here the bin size is 0.01 nW m$^{-2}$ sr$^{-1}$. Each histogram is fitted with Gaussian function shown in red solid line.}%
		\label{F13}
	\end{figure}

	Here, we used the IRTS nominal position to estimate the ISL. However, the IRTS attitude information has 1$\sigma$ error (i.e. 2.2$\arcmin$). This generates the brightness error on the IRTS ISL. To estimate the ISL error, we performed a simulation. That is, we reconstructed ISL maps in the same manner as shown above but using the shifted IRTS field positions within the IRTS astrometry error range. The shift amount was randomly chosen from the normal distribution having 1$\sigma$ of astrometry error and added to each of the IRTS nominal position. We repeated the simulation 100 times and made 100 ISL maps. The number of simulation (i.e. 100) was determined as follows. At each ISL map, we took median value of brightness. We repeated the simulation until the brightness distribution of median values shows Gaussian distribution as shown in figure \ref{F13}. Here, we used median method as each ISL map has skewed brightness distribution toward bright region due to bright Galactic stars. Among the simulated 100 ISL maps, we selected two maps and took brightness difference between pixels of them. Then, we calculated the 1$\sigma$ from the distribution of the brightness difference. This has been done for 4950 times by choosing two maps among 100 ISL maps. The 4950 is all possible combinations for choosing two maps among 100 ISL maps. The 1$\sigma$ values from the 4950 times of simulations are fairy consistent (i.e. variability relative to average is around 4\%). We defined the maximum value among 4950 simulations as ISL error which are 1.31 nW m$^{-2}$ sr$^{-1}$ and 0.71 nW m$^{-2}$ sr$^{-1}$ for 1.6 and 2.2 $\micron$, respectively.
		
	\subsubsection{Zodiacal Light}\label{SSS:irtszl}
	The ZL at near-IR consists of scattered Sun light from Interplanetary Dust (IPD) particles in the Solar system. Therefore, the ZL spectrum resembles the spectrum of the Sun. However, the ZL brightness varies with time due to the orbital motion of the Earth around the Sun (i.e. seasonal variation). \citet{kelsall98} constructed a model to estimate the ZL accounting the seasonal variation. Using this model, we reconstructed the ZL map for the IRTS field. Nevertheless, we need brightness correction of the ZL model for two reasons. First, the Kelsall model does not provide the 1.6 $\micron$ ZL brightness. Therefore, we initially obtained the 1.25 $\micron$ ZL brightness and derived the 1.6 $\micron$ ZL brightness assuming that the spectral shape does not depend on the sky position \citep{tsumura10}. Second, their exists systematic uncertainty between the ZL model and the observed ZL. The correction has been made in the same manner as \citet{matsumoto15}. The detailed descriptions are as follows.
	
\begin{figure}
		\includegraphics[width=160mm]{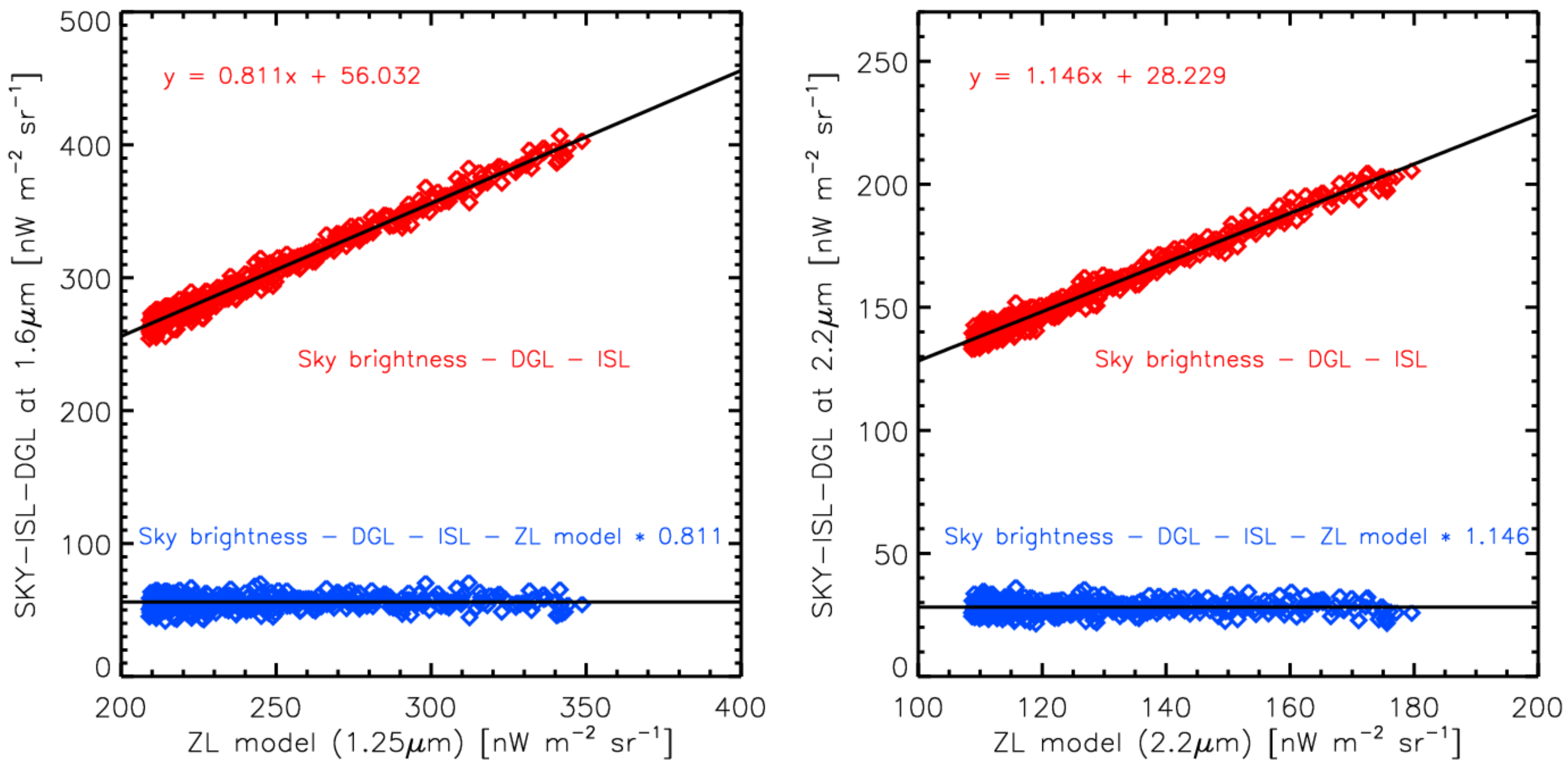}
		\caption{%
		Upper data points show correlation study between the ZL model brightness and the observed surface brightness after subtracting the ISL and DGL from the observed sky brightness. Lower data points are NIREBL brightness after subtracting the corrected ZL from the y-axis. Black solid lines are best fit lines. Left and right panels are 1.6 and 2.2 $\micron$, respectively.}%
		\label{zlcorr}
	\end{figure}

	We obtained the 1.25 and 2.2 $\micron$ ZL model brightness (hereafter, IRTS ZL) at IRTS positions using their observation date information. Then, we pixelized the IRTS ZL into the HEALPix scheme. Next, we made brightness correlation studies between the pixelized IRTS ZL model and the IRTS data after subtracting the IRTS DGL and the IRTS ISL from the IRTS SKY. Since we subtracted the DGL and the ISL, the brightness of the IRTS data contains only brightness of the actual ZL and the NIREBL. Under the assumption that the NIREBL is homogeneous and isotropic, we expect the strong correlation with the IRTS ZL model. As shown in figure \ref{zlcorr}, they show excellent correlation for both bands, which implies three things. First, the IRTS data after subtracting the IRTS DGL and the IRTS ISL from the IRTS SKY is well represented by the ZL. Second, the ZL spectral shape is uniform for the IRTS fields. Third, the NIREBL is homogeneous and isotropic emission. According to the third implication, a slope of the correlation study should show unity. However, because of band difference and systematic uncertainty between the ZL model and actual ZL, the slopes of 1.6 and 2.2 $\micron$ show 0.811 and 1.146, respectively. The band difference and systematic uncertainty were corrected by multiplying the slopes to the ZL model brightness. Then, the IRTS ZL was pixelized into the HEALPix scheme.
	
	\section{POWER SPECTRUM ANALYSIS}\label{S:irtsps}
	
	In this section, we describe the power spectrum analysis of the NIREBL. The power spectrum of the NIREBL was estimated from the pixelized NIREBL brightness map. The NIREBL brightness map was derived by subtracting the IRTS DGL, IRTS ISL and the IRTS ZL from the IRTS SKY in HEALPix format. The detailed procedure for the power spectrum analysis is described as follows.
	
	The power spectrum is an expression of the relative brightness distributions as a function of the angular scales in degree ($\theta$) or multipole moments ($\textit{l}$) related by $\theta = 180 / \textit{l}$. The relative brightness of each map is
	
	\begin{equation}
	\delta I(\Theta,\Phi)=I(\Theta,\Phi)-<I(\Theta,\Phi)>,
	\label{irtseqa1}
	\end{equation}
	
	\noindent
	where ($\Theta$, $\Phi$) are angular coordinates in the sky, and $<I(\Theta,\Phi)>$ denotes the averaged brightness of the observed region. The equation (\ref{irtseqa1}) can be decomposed into spherical harmonics as
	
	\begin{equation}
	\delta I(\Theta,\Phi)=\sum _{\textit{l}=0}^{\infty}\sum _{m=-\textit{l}}^{\textit{l}}a_{lm}Y_{lm}(\Theta,\Phi),
	\label{irtseqa2}
	\end{equation}
	
	\noindent
	where
	
	\begin{equation}
	a_{lm} \sim \Omega_{pixel} \sum _{\textit{i}=0}^{Npix} \delta I(\Theta_{i},\Phi_{i})Y^{*}_{lm}(\Theta_{i},\Phi_{i}).
	\label{irtseqa3}
	\end{equation}
	
	\noindent
	Here, $Y_{lm}(\Theta,\Phi)$ is Laplace's spherical harmonics, $a_{lm}$ is multipole coefficients of the expansion, and $\Omega_{pixel}$ is the solid angle of the HEAlPix pixel. $\textit{l}$ $=$ 0 is a monopole and $\textit{l}$ $=$ 1 is a dipole term. The power spectrum is then expressed by the variance in $a_{lm}$ as
	
	\begin{equation}
	C_{\textit{l}}=\frac{1}{2{\textit{l}}+1} \sum _{m=-\textit{l}}^{\textit{l}} \left | a_{lm} \right |^{2}.
	\label{irtseqa4}
	\end{equation}
	
	\noindent
	However, the above procedure is only valid for the full sky coverage data with no mask. If we apply it to partial sky coverage data, it produces a biased power spectrum. Therefore, we need another approach to correct for the biases since the IRTS observed only 1\% of the whole sky. There are two popular methods to measure the power spectrum for an incomplete sky coverage: maximum likelihood estimation \citep{bond98,tegmark97} and pseudo power spectrum estimation \citep{hivon02}. For the IRTS power spectrum analysis, we used the publicly available PolSpice software\footnote{http://www2.iap.fr/users/hivon/software/PolSpice/} based on the pseudo power spectrum analysis \citep{chon04}. Here, \textit{Pseudo} means that the isotropy assumption is broken and the PolSpice corrects for partial sky map.
	
	To measure the true power spectrum, PolSpice calculates $a_{lm}$ from $\delta I(\Theta,\Phi)$ map. Then, the pseudo power spectrum for incomplete sky coverage is calculated from equation (\ref{irtseqa4}). Using Legendre polynomials $P_{\textit{l}} (\cos\eta)$, the pseudo power spectrum $C_{\textit{l}}$ is then converted to correlation function
	
	\begin{equation}
	\xi(\eta)=\frac{1}{4\pi} \sum _{\textit{l}=0}^{\infty} (2\textit{l}+1) C_{\textit{l}} P_{\textit{l}} (\cos\eta),
	\label{irtseqa5}
	\end{equation}
	
	\noindent
	where $\xi(\eta)$ is the two-point correlation function defined by
	
	\begin{equation}
	\xi(\eta)= <\delta I(\Theta,\Phi) \delta I(\Theta',\Phi')>.
	\label{irtseqa6}
	\end{equation}
	
	\noindent
	Here, the angle bracket denotes the ensemble average and $\eta$ is angle between $(\Theta,\Phi)$ and $(\Theta^\prime,\Phi^\prime)$. To correct for the incomplete sky coverage, the correlation function is divided by the mask correlation function, which is estimated from the mask map where the pixel value is 0 for uncovered sky and 1 for covered sky. The corrected correlation function $\xi(\eta)$ is then inserted into the following equation to derive the true power spectrum.
	
	\begin{equation}
	C_{\textit{l}}=2\pi \int_{-1}^{1} \xi(\eta)  P_{\textit{l}}(\cos\eta) d(\cos\eta).
	\label{irtseqa7}
	\end{equation}
	
\begin{figure}
		\includegraphics[width=160mm]{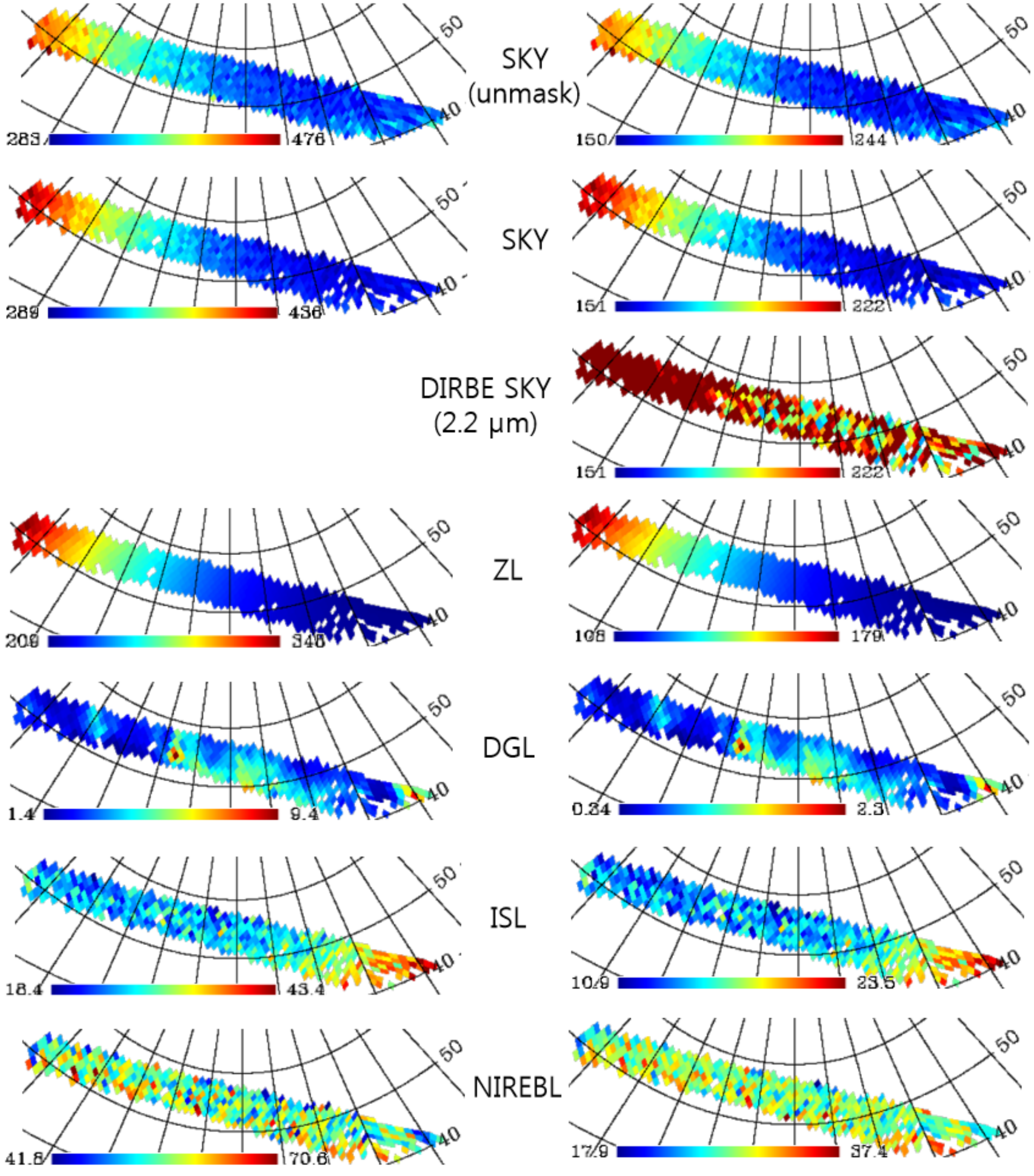}
		\caption{%
		Brightness maps for 1.6 and 2.2 $\micron$ in Galactic coordinates. Left and right maps are for 1.6 and 2.2 $\micron$, respectively. Brightness maps of IRTS raw data without mask, and IRTS raw data with mask, DIRBE 2.2 $\micron$ sky map at IRTS field, ZL, DGL, ISL, and NIREBL with mask are shown from top to bottom. Units in color bars are nW m$^{-2}$ sr$^{-1}$. DIRBE 2.2 $\micron$ sky map is shown to compare with the IRTS SKY.}%
		\label{F4}
	\end{figure}

	\noindent
	As described above, the tool needs two maps in the HEALPix format. One is a brightness map and the other is a mask map having 0 for uncovered sky and 1 for covered sky. Using the NIREBL brightness map in figure \ref{F4}, we calculated the power spectrum. However, the NIREBL power spectrum still contains photon and readout noise. Since the noise level for each IRTS SKY is unknown, we performed a Monte Carlo simulation to estimate the noise power spectrum. First, we made a Gaussian distribution having a 1$\sigma$ of readout noise and photon noise \citep{matsumoto05}. Under the Gaussian distribution assumption, we randomly picked the value in the distribution and assigned it to each IRTS field. The noise assigned map is then pixelized into the HEALPix scheme. Then, the power spectrum was estimated for the pixelized noise map. This procedure was repeated 100 times and averaged to derive the final noise power spectrum. The noise power spectrum was then subtracted from the NIREBL power spectrum for 1.6 and 2.2 $\micron$ of the IRTS.
	
	Nevertheless, the finite resolution and pixelization can suppress the power spectrum at a small angular scale. The suppression can be corrected using a beam transfer function that depends on the shape and size of the PSF. However, we concluded that the beam transfer function for the IRTS/NIRS beam does not affect the fluctuation spectrum above 2$^{\circ}$, which was twice the HEALPix pixel resolution. Since the power spectrum was only valid for angular scales above 2$^{\circ}$ according to the Nyquist sampling, we did not apply the correction for the beam transfer function.

	\section{ERROR ESTIMATION}\label{S:error}
	
	Errors can be categorized into random and systematic components. Random errors include sample variance of the power spectrum (i.e. \textit{$\delta I_{variance}$}), attitude error of the IRTS (i.e. \textit{$\delta I_{attitude}$}), model error of the DGL (i.e. \textit{$\delta I_{DGL}$}), and binning error of the HEALPix pixel (i.e. \textit{$\delta I_{binning}$}). \textit{$\delta I_{variance}$} is the error induced from the power spectrum estimation. At a given angular scale, the number of possible modes is limited due to a finite sky coverage. The smaller angular scale has smaller sample variance due to larger number of possible modes. However, we cannot estimate the error directly from the power spectrum analysis since the small coverage of the IRTS field results in the covariance matrix of the sample variance being very noisy. Alternatively, we used the empirically determined Knox formula that represents the $\chi^2$ distribution of $C_{\textit{l}}$ with its mean \citep{knox95}. The formula is given in Appendix C of \citet{thacker15}. \textit{$\delta I_{attitude}$} is the error induced from inaccurate IRTS attitude. We described the detailed procedure of \textit{$\delta I_{attitude}$} calculation in section \ref{SSS:irtsisl}. \textit{$\delta I_{DGL}$} is transferred from the scale factor which is needed to convert 100 $\micron$ intensity (i.e. far-IR) to near-IR DGL brightness as described in section \ref{SSS:irtsdgl}. \citet{arai15} compared the DGL model to the observed data and found that the scale factor has 20\% uncertainty. To account this error, we multiplied 1.2 to the scale factor and derived the DGL brightness ${I\textprime}_{DGL}$. Then, we subtracted nominal DGL brightness $I_{DGL}$ from ${I\textprime}_{DGL}$. Finally, \textit{$\delta I_{DGL}$} was computed by taking 1$\sigma$ of the brightness difference (i.e. ${I\textprime}_{DGL} - {I_{DGL}}$). \textit{$\delta I_{binning}$} is generated when computing mean of the IRTS pixel brightness belong to each HEALPix pixel. Our procedure to calculate \textit{$\delta I_{binning}$} is as follows. First, we made binning error map by taking 1$\sigma$ of IRTS pixel brightness belong to each HEALPix pixel. Then, computed 1$\sigma$ of the binning error map. This procedure was done for each astrophysical components as well as the sky. Finally, \textit{$\delta I_{binning}$} was calculated by combining all error components in a quadrature.
		
	\begin{equation}
	\delta I_{binning} = \sqrt{\delta I_{SKY,binning}^{2}+\delta I_{DGL,binning}^{2}+\delta I_{ISL,binning}^{2}+\delta I_{ZL,binning}^{2}}
	\label{irtseqa8}
	\end{equation}
	
	\noindent
	Then, we combined all random errors using following equation excepting \textit{$\delta I_{variance}$} where the power spectrum of \textit{$\delta I_{variance}$} was directly calculated using the Knox formula \citep{knox95}.
	
	\begin{equation}
	\delta {I\textprime}_{random} = \sqrt{\delta I_{attitude}^{2}+\delta I_{DGL}^{2}+\delta I_{binning}^{2}}
	\label{irtseqa9}
	\end{equation}
	
	\noindent
	The power spectrum of the \textit{$\delta {I\textprime}_{random}$} was calculated based on the Monte Carlo simulation in the same manner as power spectrum calculation for the readout and photon noises described in this section. Then, power spectrum of total random error \textit{$\delta {C}_{{l, random}}$} was calculated using following equation.
	
	\begin{equation}
	\delta {C}_{{l, random}} = \sqrt{{\delta C\textprime}_{{l, random}}^{2}+\delta {C}_{{l, variance}}^{2}}
	\label{irtseqa10}
	\end{equation}
	
	\noindent
	where ${\delta C\textprime}_{l, random}$ is power spectrum of $\delta {I\textprime}_{random}$ and $\delta {C}_{{l, variance}}$ is power spectrum of \textit{$\delta I_{variance}$}.
	
	The systematic errors are categorized into two. One is the calibration error of the IRTS (i.e. \textit{$\delta I_{cal}$}) which is known as 3\% of the NIREBL brightness \citep{matsumoto05,matsumoto15}. The other is the limiting magnitude error of the IRTS (i.e. \textit{$\delta I_{lim}$}) where the IRTS has $\pm$ 0.5 limiting magnitude uncertainty (see figure \ref{F3}). To compute \textit{$\delta I_{cal}$}, we multiplied the 0.03 to the NIREBL brightness map then took 1$\sigma$ of it. \textit{$\delta I_{lim}$} causes the systematic brightness difference on the ISL. To compute this, we derived the ISL brightness error (i.e. \textit{$\delta {I\textprime}_{lim}$}) after adding $\pm$ 0.5 to the limiting magnitude of the IRTS. The \textit{$\delta {I\textprime}_{lim}$} derivation was same as the nominal ISL brightness (i.e. \textit{$\delta {I}_{nominal}$}) derivation described in section \ref{SSS:irtsisl}. Then, we took brightness difference between \textit{$\delta {I\textprime}_{lim}$} and \textit{$\delta {I}_{nominal}$}. The \textit{$\delta {I}_{lim}$} was then calculated by taking 1$\sigma$ of the brightness difference. Finally, total systematic error was derived as
	
	\begin{equation}
	\delta {I}_{systematic} = \sqrt{\delta I_{cal}^{2}+\delta I_{lim}^{2}}
	\label{irtseqa11}
	\end{equation}
	
	\noindent 
	Then, the power spectrum of the \textit{$\delta {I}_{systematic}$} was calculated based on the Monte Carlo simulation in the same manner as $\delta {C}_{{l, random}}$. Finally, the power spectrum of total error was calculated using following equation.
	
	\begin{equation}
	\delta {C}_{{l, total}} = \sqrt{{\delta C}_{{l, random}}^{2}+\delta {C}_{{l, systematic}}^{2}}
	\label{irtseqa10}
	\end{equation}
	
	\noindent
	where $\delta {C}_{{l, systematic}}$ is power spectrum of \textit{$\delta {I}_{systematic}$}.	The brightness of errors listed above excepting \textit{$\delta I_{variance}$} are shown in Table \ref{T1}.

	\section{RESULT}\label{S:result}
	
	The fluctuation spectra of the NIREBL for 1.6 and 2.2 $\micron$ are shown in figures \ref{F8} and \ref{F9}, respectively. In figure \ref{F8}, the IRTS fluctuation shows power-law with angular scale. The power index close to -1 indicates that the fluctuation is random and structureless. This homogeneous and isotropic feature is also shown in the NIREBL map in figure \ref{F4}. For comparison, the fluctuation spectrum of the CIBER project \citep{zemcov14} at smaller angular scale ($<$ 1$^{\circ}$) was also shown. It seems to have smooth connection between the IRTS and the CIBER. Although there exists no data between them, we can infer that the NIREBL fluctuation has a peak at around 1$^{\circ}$.
	
	In figure \ref{F9}, we also examined the fluctuation at the 2.2 $\micron$. Since the CIBER does not have a 2.2 $\micron$ data, we needed to multiply a scale factor to the CIBER 1.6 $\micron$ power spectrum. The scale factor was derived by the IRTS 1.6/2.2 $\micron$ color ratio assuming color of the NIREBL does not depend on the sky position. The ratio was derived from the IRTS 1.6 and 2.2 $\micron$ correlation study as shown in figure \ref{F10}. Then the ratio was multiplied to the CIBER 1.6 $\micron$ power spectrum to derive the CIBER 2.2 $\micron$ one. As well as the CIBER, we also compared fluctuation spectra for AKARI 2.4 $\micron$ \citep{matsumoto11,seo15} and Spitzer 3.6 $\micron$ \citep{kashlinsky12} by scaling the amplitude of their fluctuations to the 2.2 $\micron$ under the Rayleigh-Jeans assumption \citep{matsumoto11}. 

	As a result, the fluctuation spectra from the AKARI and the Spitzer are marginally consistent with each other. Especially, the Spitzer is extended toward larger angular scales where the fluctuation amplitude is 10 times larger than that of the ILG. The discrepancy between the CIBER and the other measurements (i.e. Spitzer and AKARI) may indicate that the contributing components of the NIREBL are somehow different depending on the wavelengths. Otherwise, since the scaling is only valid if fluctuation follows Rayleigh-Jeans like spectrum, the discrepancy may occurred by the scale factors. Nevertheless, their similar spectral shapes indicate they have same origin. Furthermore, the fluctuation spectrum of the IRTS at 2$^{\circ}$ is located at the middle of the CIBER and the Spitzer fluctuations so either case implies a peak at around 1$^{\circ}$ angular scale. Evidently, the 1$^{\circ}$ peak seems to be a common feature in the broad wavelength ranges at near-IR.
	
	We also examined the IRTS 1.6 and 2.2 $\micron$ brightness correlation. As shown in figure \ref{F10}, the 1.6 and 2.2 $\micron$ show excellent correlations. The 1.6/2.2 $\micron$ ratio is consistent with that from the NIREBL brightness spectrum obtained by \citet{matsumoto15}. In addition, we derived the absolute brightness of the NIREBL using result shown in figure \ref{zlcorr}. The y-intercept where the brightness of the ZL becomes zero represents the absolute brightness of the NIREBL. They are 56.032 and 28.228 nW m$^{-2}$ sr$^{-1}$ for the 1.6 and 2.2 $\micron$, respectively. They are fairly consistent with those of \citet{matsumoto15} which implies that our NIREBL brightness derivation is reasonable. This confirms consistency in the data analysis, and the excess fluctuation is strongly associated with the NIREBL spectrum.
	
	\section{DISCUSSION}\label{S:discussion}
	
	To find the possible origin of excess power at around 1$^{\circ}$, we examined several candidate sources. High redshift objects (e.g. first stars) were initially excluded since they show a turn over at around 0.3$^{\circ}$, which contradicts with the peak fluctuation at around 1$^{\circ}$ (see figure A-2 in \cite{kashlinsky12}).
	
	The first candidate is the foregrounds such as ISL, DGL, and ZL. To check this, we analized the cross correlation between each foreground component and NIREBL using the PolSpice analysis tool. However, none of the foreground components show correlation with the NIREBL. As a reference, the ISL and the DGL fluctuations are shown in Figure \ref{F26}. Since the ZL is based only on the model, we do not evaluate the ZL fluctuation in this work. According to the \citet{zemcov14}, however, the ZL fluctuation is too small to detect the fluctuation power.
	
\begin{figure}
		\includegraphics[width=160mm]{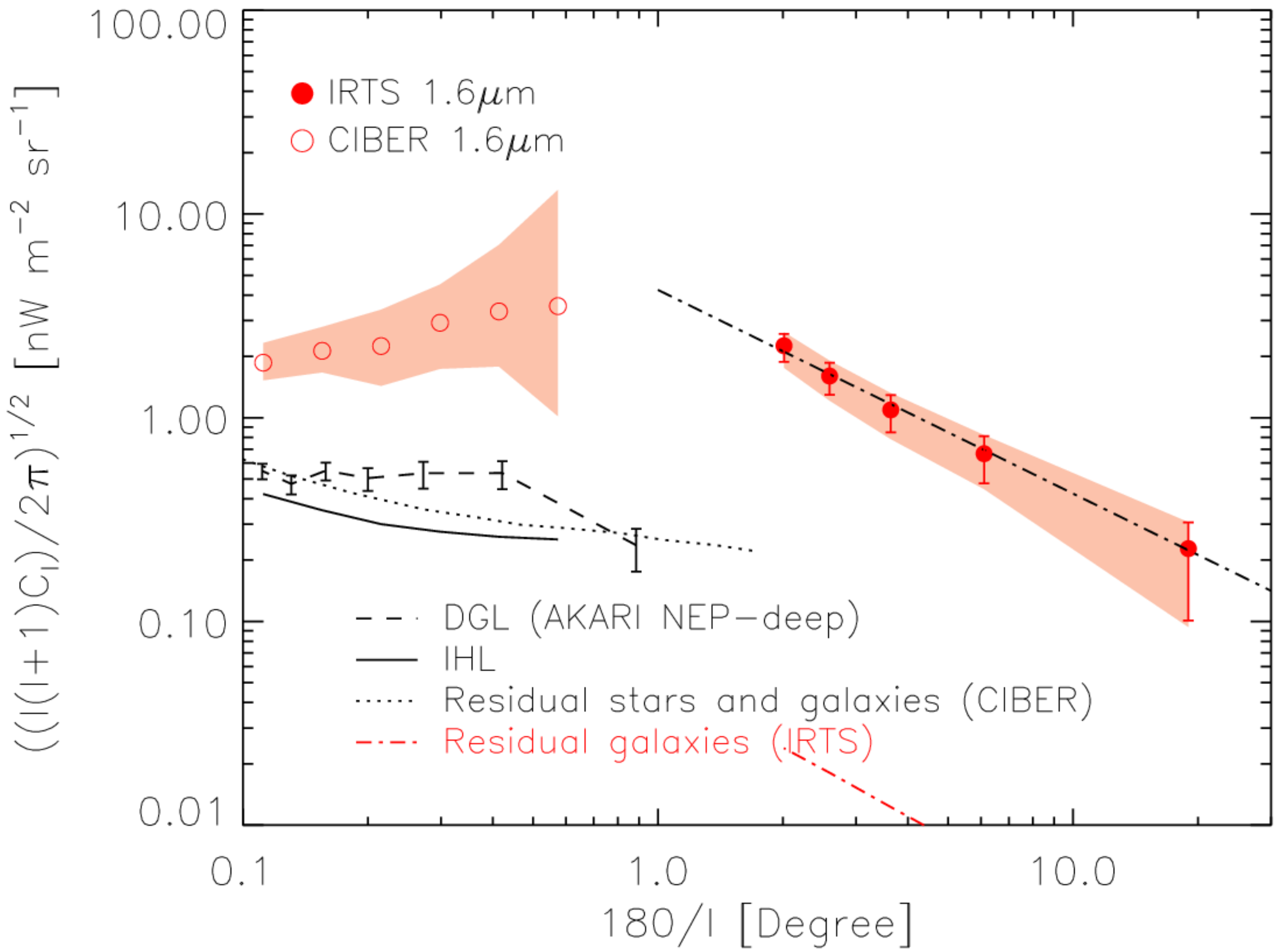}
		\caption{%
		The measured 1.6 $\micron$ fluctuations for the IRTS. The IRTS (this work) and the CIBER \citep{zemcov14} auto spectra are in filled and unfilled red circle, respectively. The shaded color of the IRTS shows the error including systematic error and the random error is drawn with error bar. The shaded color for the CIBER denotes estimated errors. Black dot-dashed line is a power-law with index -1. The red dot-dashed line is spectrum due to unresolved galaxies. Dashed line is DGL spectrum measured using the AKARI/FIS deep pointing data toward NEP region \citep{seo15}. Dotted and solid lines are unmasked sources (i.e. stars and galaxies for $m_{H}$ $>$ 17) and IHL spectrum from \citet{zemcov14}, respectively.}%
		\label{F8}
	\end{figure}

\begin{figure}
		\includegraphics[width=160mm]{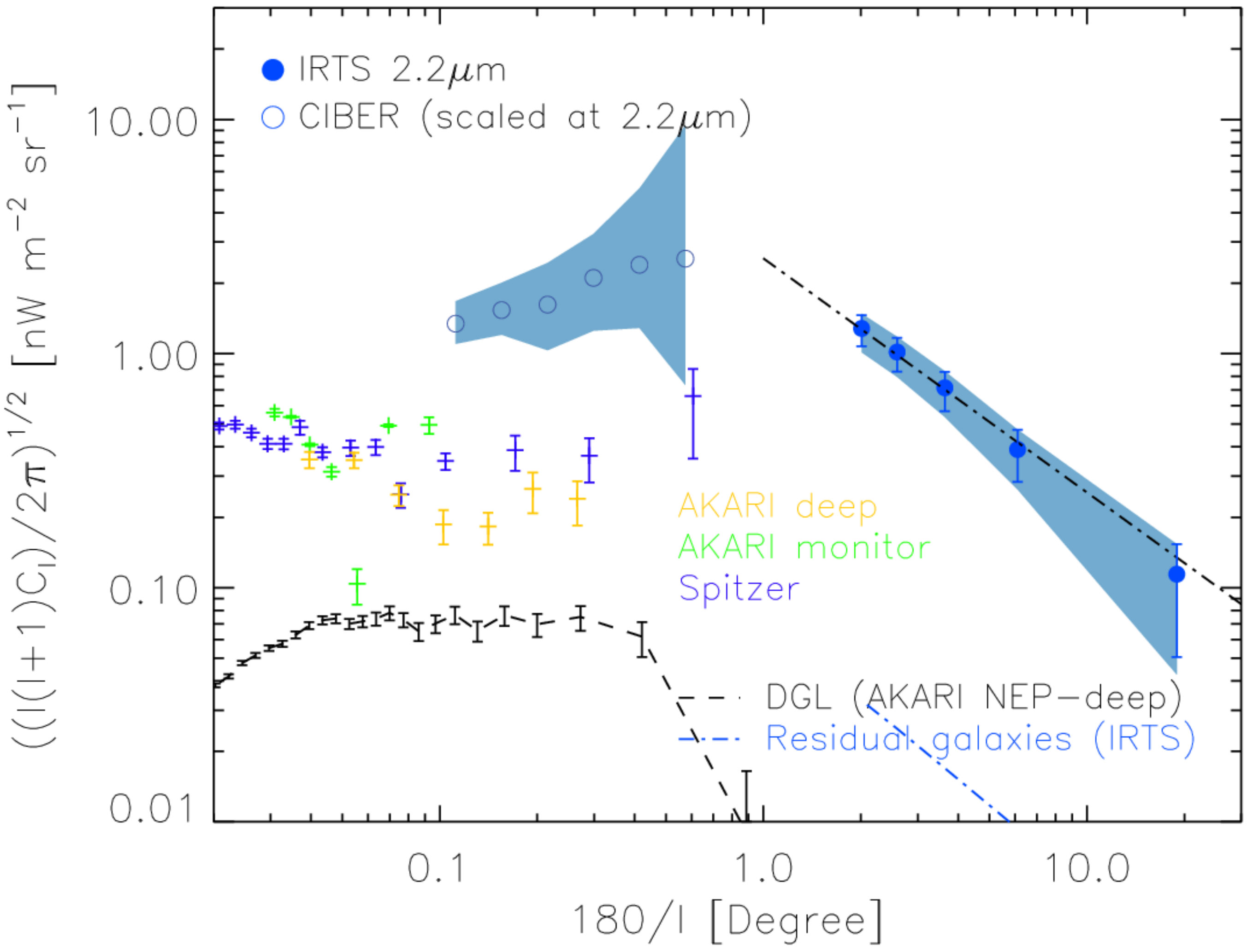}
		\caption{%
		The measured 2.2 $\micron$ fluctuations for the IRTS. The IRTS (this work) and the CIBER \citep{zemcov14} auto spectra are in filled and unfilled blue circles, respectively. The shaded color of the IRTS shows the error including systematic error and the random error is drawn with error bar. The shaded color for the CIBER denotes estimated errors. Black dot-dashed line is a power-law with index -1. The blue dot-dashed line is spectrum due to unresolved galaxies. The CIBER 1.6 $\micron$ is scaled to 2.2 $\micron$ using IRTS 1.6/2.2 $\micron$ color ratio. Plus signs with errors are the fluctuation spectra from the AKARI 2.4 $\micron$ \citep{matsumoto11,seo15} and Spitzer 3.6 $\micron$ \citep{kashlinsky12}. Their spectra are scaled to 2.2 $\micron$ under the Rayleigh-Jeans assumption. Dashed line is DGL spectrum measured using the AKARI/FIS deep pointing data toward NEP region \citep{seo15}.}%
		\label{F9}
	\end{figure}

\begin{figure}
		\includegraphics[width=160mm]{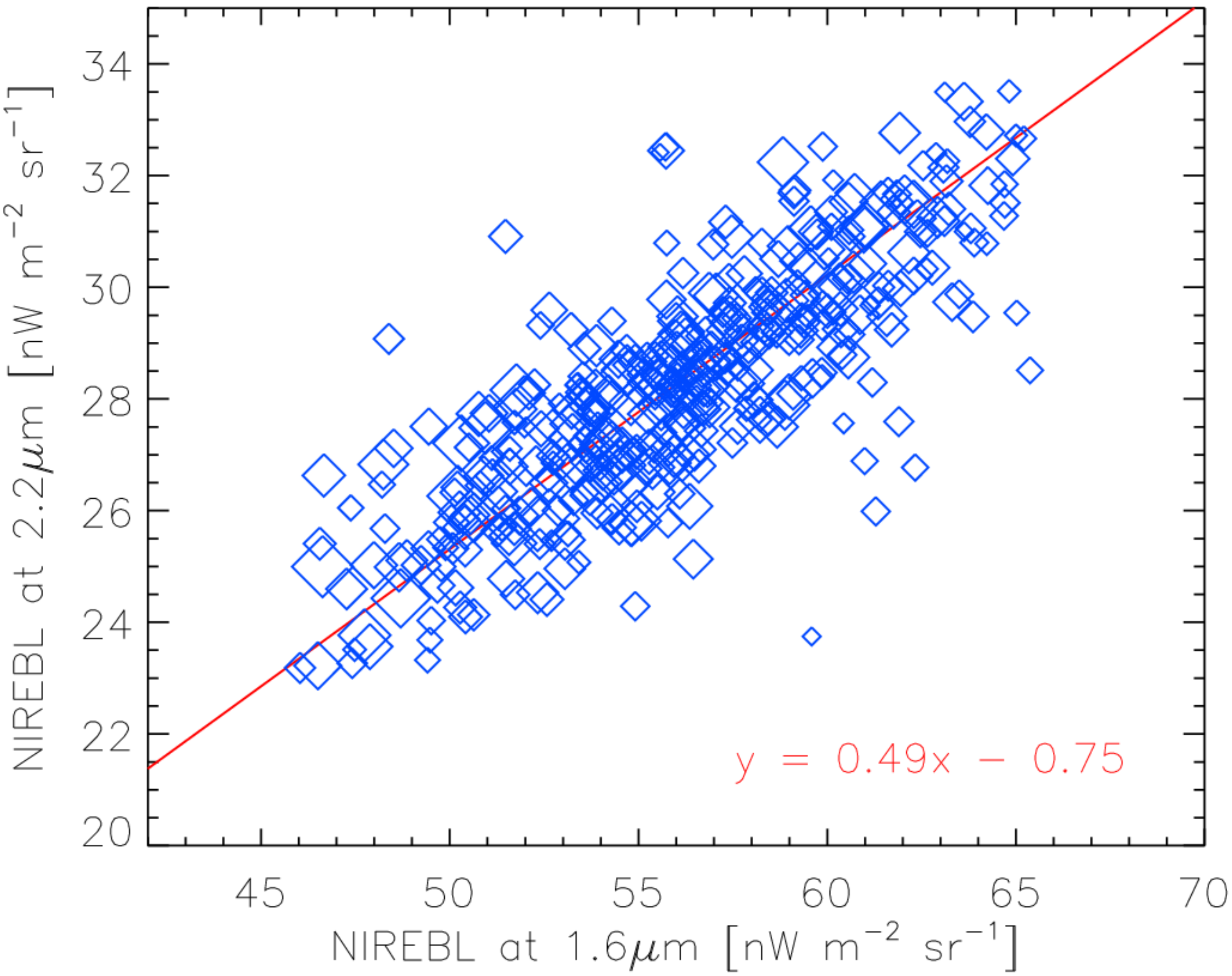}
		\caption{%
		The correlation between the 1.6 and 2.2 $\micron$ NIREBL brightness after subtraction of the astrophysical foreground components from the IRTS data. Each data point has different symbol size inversely weighted by its error. That is, the lager symbol represents the smaller error. The best linear fit is shown in red solid line. The fitting parameters are shown in the bottom-right corner of the figure.}%
		\label{F10}
	\end{figure}

\begin{figure}
		\includegraphics[width=160mm]{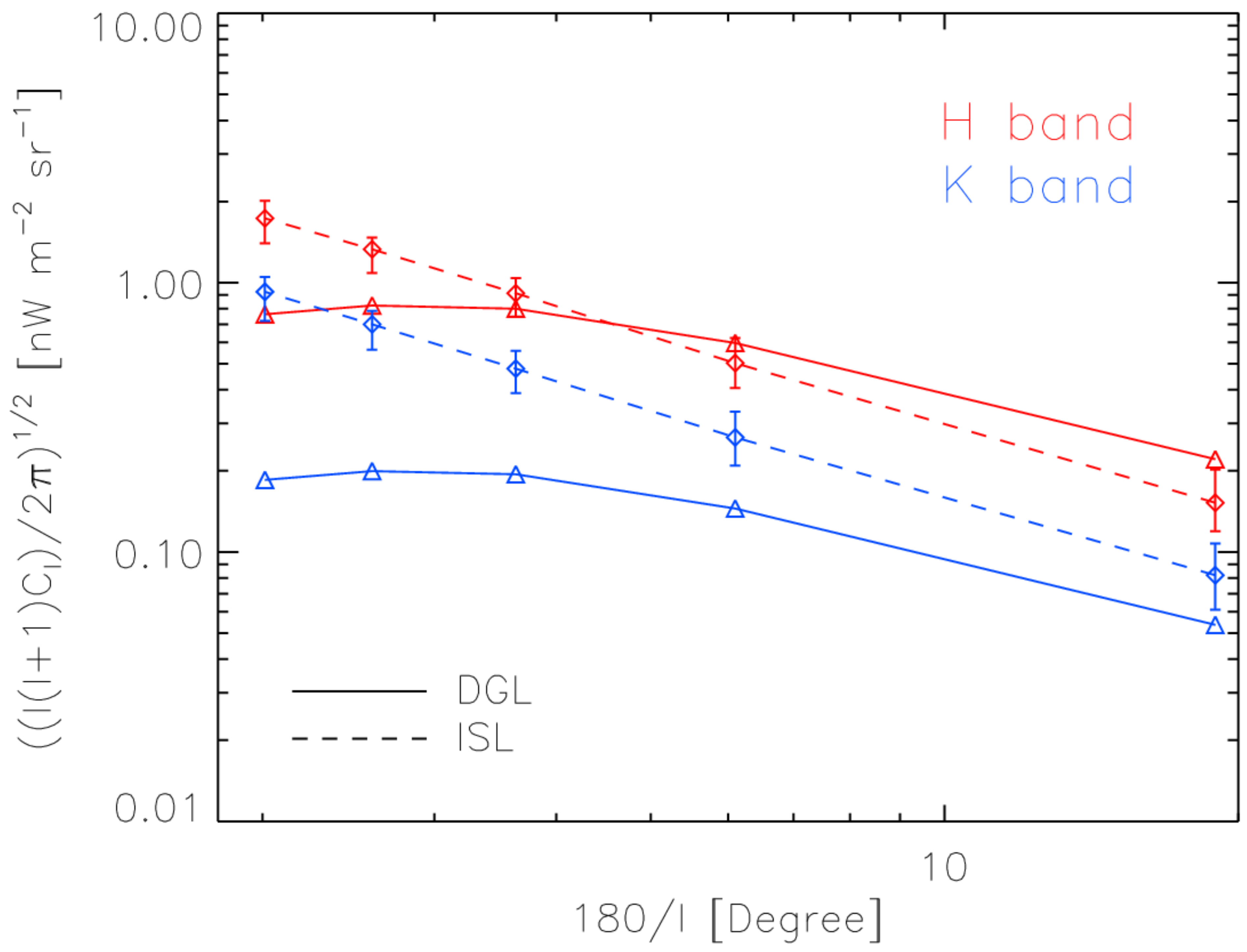}
		\caption{%
			The foreground fluctuations of the IRTS fields. The ZL fluctuation is not shown since the ZL is based only on the model and expected to have very small fluctuation \citep{zemcov14}.}%
		\label{F26}
	\end{figure}

	The second candidate is the IHL. The IHL fluctuation can be composed of one-halo and two-halo terms \citep{cooray12a}. The one-halo term describes the clustering of baryonic matter inside a halo, and the two-halo term describes the correlations between the individual halos. The two-halo term shows larger power fluctuation than the one-halo term. In figure \ref{F8}, we drew the contribution of the IHL from \citet{zemcov14} and compared it to the NIREBL fluctuations. Although the IHL spectrum was only estimated at sub-degree scales, the amplitude was too low to explain the excess at 1$^{\circ}$.
	
	Next one is the DGL which accounts for a large portion of the fluctuation at a small angular scale measurement \citep{zemcov14}. \citet{gautier92} empirically derived that the DGL power spectrum (\textit{$C_l$}) follows \textit{$l^{-3}$} of power-law. If we extend the DGL spectrum at \citet{zemcov14} toward a larger angular scale with the constant power-law, the excess emission of the IRTS can be explained. However, their DGL estimation was obtained from low angular resolution map \citep{schlegel98} and we may expect a slower increase than $\theta^3$ towards larger angular scales.
	
	To measure the DGL directly without a power-law extrapolation at sub-degree scales, we used a high resolution (pixel scale $\sim$ 0.16$\arcmin$) and deep pointing AKARI 90 $\micron$ image of the North Ecliptic Pole (NEP) region \citep{seo15}. The intensity of the map was scaled to near-IR using the empirical scaling relation in the same manner as described in section \ref{SSS:irtsdgl}. We then measured the power spectrum using POKER\footnote{http://www.ipag.osug.fr/$\sim$ponthien/Poker/Poker.html} which is a publicly available tool. The POKER estimates the power spectrum using a Fourier transform under the flat sky approximation. Since projecting the observed sky onto a plane distorts the data, it is only applicable for an image of less than a few degrees \citep{pont11}. The flat sky approximation is valid for the AKARI image with a FoV of 1.2$^{\circ}$ $\times$ 1.2$^{\circ}$. 
	
	The estimated DGL fluctuation is consistent with the CIBER at smaller angular scales but decreases toward larger angular scales as shown in figures \ref{F8} and \ref{F9}. Although the DGL intensity depends on the field to field, the overall shape of the DGL fluctuation at degree scale also decreases as shown in figure \ref{F26}. Furthermore, the cross correlation between the DGL and the NIREBL indicates that the DGL does not contribute to the NIREBL.
	
\begin{figure}
		\includegraphics[width=160mm]{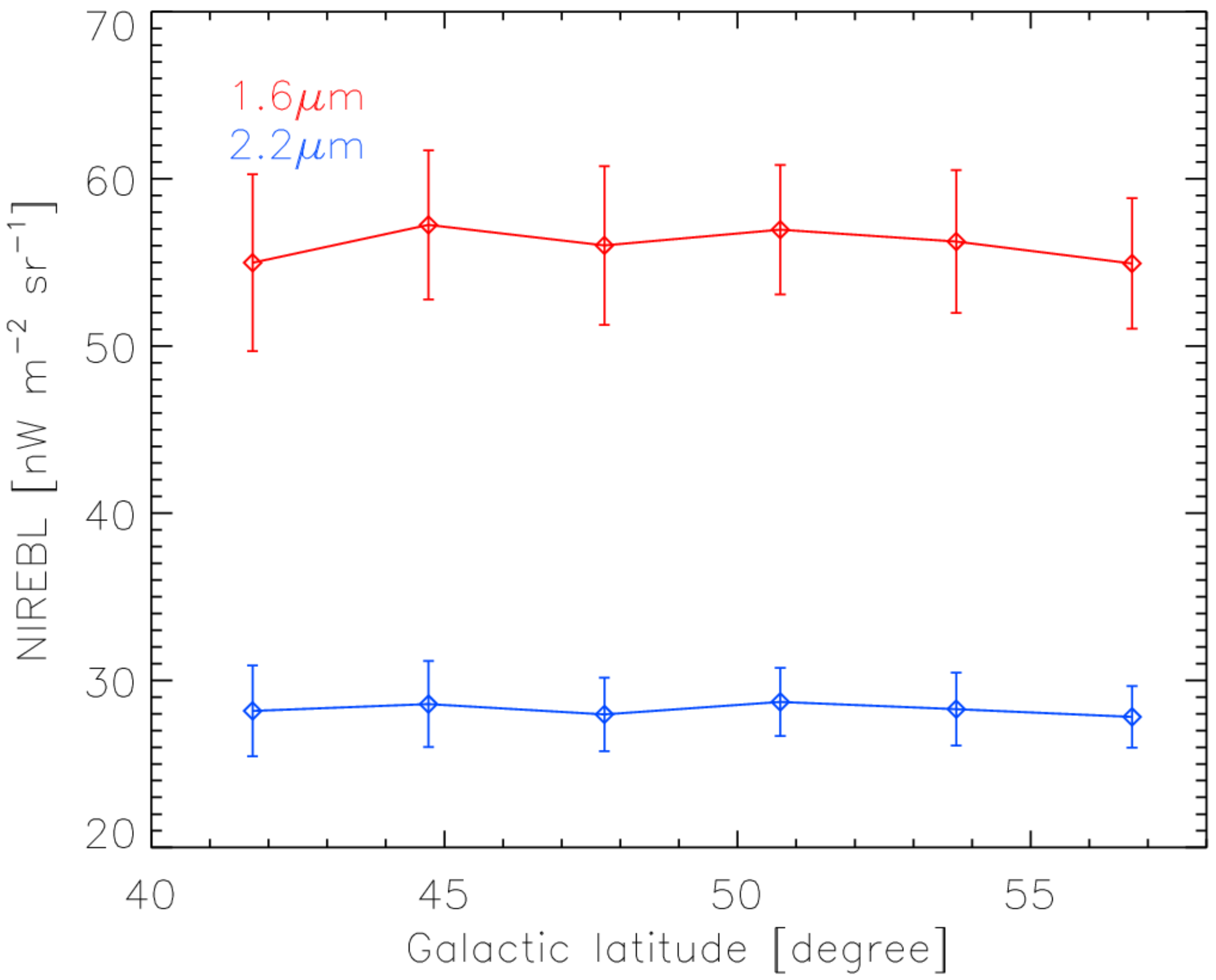}
		\caption{%
		The NIREBL brightness dependence along the Galactic latitude bin. Here the bin size is 3$^{\circ}$. The 1 $\sigma$ error of each bin is drawn. The upper and lower curves are for 1.6 and 2.2 $\micron$, respectively.}%
		\label{isldep}
	\end{figure}

	We examined the possibility of stellar contamination on the residual background, which can be imperfectly subtracted. If the NIREBL has residual stellar contribution, the 1.6/2.2 $\micron$ ratios of the NIREBL and that of the stars are similar. To check this, we derived 1.6/2.2 $\micron$ ratio of 2MASS stars in the IRTS fields. The derived ratio is 0.57 which is only 15\% steeper than the 1.6/2.2 $\micron$ ratio of the NIREBL in figure \ref{F10}. Since the difference is not significant, we additionally checked whether a Galactic latitude dependency exists on the NIREBL map as shown in figure \ref{isldep}. Here, we averaged the brightness of the NIREBL map with constant interval (i.e. 3$^{\circ}$) of the Galactic latitude. Nevertheless, it shows no dependency along the Galactic latitude, which indicates that the Galactic stars are not proper candidates.
	
	We compared the IRTS with the DIRBE (see figure \ref{F4} for DIRBE map). Since the DIRBE has much brighter detection limit (i.e. $\sim$3 mag), stellar contribution is mainly due to nearby bright stars and thus no Galactic latitude dependency is shown. However, they carefully subtracted the Galactic stars to achieve homogeneous background map. Using much more sensitive IRTS image but poor attitude information, the 1$\sigma$ of the NIREBL brightness distribution at 2.2 $\micron$ is 2.16 nW m$^{-2}$ sr$^{-1}$ which is fairy consistent with the DIRBE 2.2 $\micron$ study \citep{levenson07}.
	
	A fraction of the NIREBL brightness is also contributed by normal galaxies. To estimate contribution of the galaxy at the degree scales, we performed Monte Carlo simulations using the galaxy count model presented by \citet{keenan10}. We made a brightness map by randomly distributing the galaxies in the sky based on the model count and calculated the power spectrum. We then repeated this procedure 10 times and made 10 maps. We calculated the power spectrum of each map and took average of those power spectra. Nevertheless, they contribute less than 1\% of the NIREBL fluctuation level for 1.6 and 2.2 $\micron$ at large angular scales (see figures \ref{F8} and \ref{F9}).
	
\begin{figure}
		\includegraphics[width=160mm]{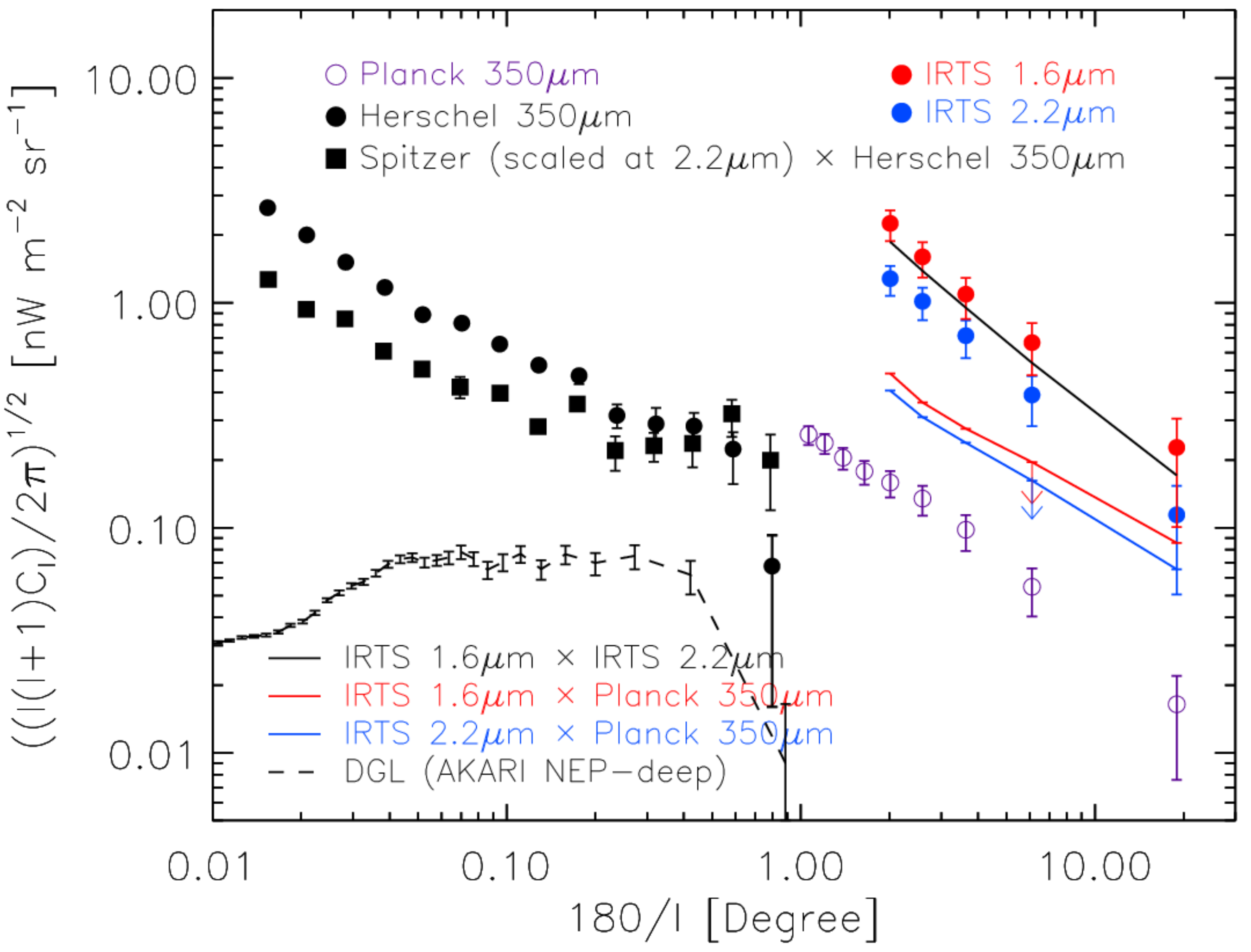}
		\caption{%
		Auto correlation fluctuation spectra of Herschel 350 $\micron$ \citep{thacker15} and Planck 350 $\micron$ (this work), together with cross correlations of Spitzer 3.6 $\micron$ and Herschel 350 $\micron$ \citep{thacker15}, IRTS 1.6 $\micron$ and Planck 350 $\micron$ (this work), and IRTS 2.2 $\micron$ and Planck 350 $\micron$ (this work). Auto correlation fluctuation spectra of IRTS 1.6 $\micron$ and 2.2 $\micron$ (this work) are also drawn. Only upper limits (arrows with solid lines) of cross correlation between the IRTS bands and Planck are shown because of large errors. The Spitzer spectrum is scaled to 2.2 $\micron$ under the Rayleigh-Jeans assumption. Dashed line is DGL spectrum measured using the AKARI/FIS deep pointing data toward NEP region \citep{seo15}.}%
		\label{F11}
	\end{figure}

	We also examined the near-IR and far-IR cross correlation using the PolSpice analysis tool. For the near-IR map, we used the NIREBL map reduced from this work. For the far-IR map, we used the Cosmic Infrared Background (CIB) map reduced from the Planck 857 GHz data. The Planck team provides a Galactic thermal dust and Cosmic Microwave Background (CMB) removed Planck map\footnote{http://pla.esac.esa.int/pla} where the residual brightness contains only the extragalactic CIB component. According to \citet{planck11}, the Planck map is composed of dusty, star forming galaxies mostly from a low redshift (z $<$ 0.8). The auto correlation spectrum for Planck is shown in figure \ref{F11}. The fluctuation seems to have smooth connection with Herschel 350 $\micron$ one \citep{thacker15} toward smaller angular scales. 
	
	Interestingly, the IRTS and the Planck show a good correlation, although only the upper bound is shown due to a large error (see figure \ref{F11}). If the nominal cross spectrum is near the upper bound, the fluctuation spectrum (i.e. IRTS K cross Planck) smoothly connects with the Spitzer (3.6 $\micron$) cross the Herschel (350 $\micron$) spectrum \citep{thacker15} having a peak at around 1$^{\circ}$ angular scale. Note that the Spitzer (3.6 $\micron$) spectrum is scaled to the IRTS 2.2 $\micron$ under the Rayleigh-Jeans assumption for the comparison. According to the measurements at sub-degree scale by \citet{thacker15}, about half of the near-IR background can be explained by dusty, star forming galaxies, and the residuals can be explained by the IHL. However, it is difficult to explain the whole excess at degree scales since only the upper limits were obtained.
	
	Since the sources contributing to the fluctuations of the near-IR background at such large angular scales have not been clearly identified, we examined possible candidates. However, none of them seem to show a significant contribution. Thus, future studies are necessary to understand the anisotropies from sub-degree to degree scales.
	
	\section{SUMMARY}\label{S:summary}
	We measure the NIREBL fluctuation spectra at angular scales between 2$^{\circ}$ to 20$^{\circ}$ for the 1.6 and 2.2 $\micron$ for the first time. The NIREBL power spectrum is calculated from the NIREBL brightness map after subtracting the foreground components, such as the DGL, the ISL, and the ZL from the observed sky brightness. The readout and photon noises of the IRTS are subtracted from the power spectrum. Within the range of the angular scale studied here, the NIREBL fluctuation monotonically declines with \textit{F($\sqrt{l(l+1)C_l/2\pi}$}) $\sim$ \textit{$\theta^{-1}$} constant power-law indicating that the fluctuations at an angular scale greater than 2$^{\circ}$ is random and structureless. The bumpy structures in \citet{matsumoto05} is also reduced in this work by correcting the effect of the mask pixels on the power spectrum. Furthermore, comparing with \citet{matsumoto05}, our study achieves larger sky coverage and thus larger 2-dimensional sampling which enables us to compare fluctuations with other studies directly. Our result also consistent with \citet{matsumoto15} for the 1.6 and 2.2 $\micron$ absolute brightness measurement. This implies that the sky fluctuation is strongly related to the NIREBL spectrum. Comparing the results with previous studies at sub-degree scales, both the 1.6 and 2.2 $\micron$ spectra appear to have broad bumps with a center at around 1$^{\circ}$. We examine several proposed origins explaining the sub-degree scale fluctuations, but these are not likely contribute to the fluctuations at degree scales. Interestingly, the fluctuations at 857 GHz with Planck after subtraction of the foreground and CMB suggest a good correlations with those of the IRTS bands, although we can only set the upper limit due to large uncertainties. If they have a significant correlation, this indicates that some portions of the anisotropies at degree scales can be explained by dusty, star forming galaxies at $z < 0.8$. Recently, the Korean space mission MIRIS performed deep observations toward the large area near the NEP region (10$^{\circ}$ $\times$ 10$^{\circ}$). The data is being processed and is expected to probe the fluctuations in the spectrum at around 1$^{\circ}$ to several degree scales. This work provides motivation to study various kinds of background that can contribute to the degree scale fluctuations.
	
	\newpage
	
	\clearpage
	
	\clearpage
	
	\clearpage
	
	\clearpage
	
	\clearpage
	
	\clearpage
	
	\clearpage
	
	\clearpage
	
	\clearpage
	
	\clearpage
	
	\clearpage
	
	\clearpage
	
	\clearpage
		
	\clearpage
	
	\clearpage

\begin{table}
	\tbl{Error budget for the NIREBL fluctuation}{%
		\begin{tabular}{|c|ccc|cc|}
			\hline
			\multicolumn{1}{|l|}{}  & \multicolumn{3}{c|}{Statistical error} & \multicolumn{2}{c|}{Systematic error} \\ \hline
			Band & ISL attitude error & DGL scale factor error & Healpix binning error & IRTS Calibration error & ISL limiting mag error           \\
			1.6 $\micron$ & 1.31 & 0.24 & 0.08 & 0.13 & 1.86 \\
			2.2 $\micron$ & 0.71 & 0.06 & 0.04 & 0.06 & 0.98 \\
			\hline
			\end{tabular}}\label{T1}
			\begin{tabnote}
			\footnotemark[$*$] The sample variance is not shown since it is estimated by empirically determined Knox formula.  \\ 
			\footnotemark[$**$] Units are nW m$^{-2}$ sr$^{-1}$.  \\ 
	\end{tabnote}
\end{table}

	\clearpage

\begin{ack}
	{\normalsize This work is based on the observations with the IRTS.
	M.G.K. acknowledges support from the Global PhD Fellowship Program through
	the NRF, funded by the Ministry of Education (2011-0007760).
	H.M.L. was supported by NRF grant 2012R1A4A1028713. K.T. was supported by JSPS KAKENHI (17K18789 and 18KK0089).
	W.-S.J. acknowledges support from the National Research Foundation of Korea (NRF) grant funded by the Ministry of Science and	ICT (MSIT) of Korea (NRF-2018M1A3A3A02065645). The authors thank Dr. Hivon for providing Polspice and Dr. Ponthieu for providing POKER power spectrum analysis tool. We also thank Dr. Girardi who enables us to model the Galactic stars using the TRILEGAL code. This publication makes use of data products from the 2MASS, which is a joint project of the University of Massachusetts	and the Infrared Processing and Analysis Center/California Institute of Technology,
	funded by the NASA and the NSF. Based on observations obtained with Planck (http://www.esa.int/Planck), an ESA science mission with instruments and contributions directly funded by ESA Member States, NASA, and Canada.}
\end{ack}

\end{document}